%% file: main_sc.tex
\RequirePackage[OT1]{fontenc}
\documentclass[journal]{IEEEtran}
\usepackage{amsmath}
\usepackage{amsfonts}
\usepackage{algorithmic}
\usepackage{algorithm}
\usepackage{array}
\usepackage[caption=false,font=normalsize,labelfont=sf,textfont=sf]{subfig}
\usepackage{textcomp}
\usepackage{stfloats}
\usepackage{url}
\usepackage{verbatim}
\usepackage{graphicx}
\usepackage{cite}
\usepackage{tikz}
\usepackage{amssymb, flushend}
\usepackage{setspace}
\usepackage{pgfplots}
\pgfplotsset{compat=newest}
\usepackage{subfig}  
\usepackage{float}
\def\BibTeX{{\rm B\kern-.05em{\sc i\kern-.025em b}\kern-.08em
    T\kern-.1667em\lower.7ex\hbox{E}\kern-.125emX}}
\AtBeginDocument{\definecolor{ojcolor}{cmyk}{0.93,0.59,0.15,0.02}}

\begin{document}

	
\title{SORIS: A Self-Organized Reconfigurable Intelligent Surface Architecture for Wireless Communications}
\author{Evangelos Koutsonas, \IEEEmembership{Student~Member, IEEE}, Alexandros-Apostolos A. Boulogeorgos,  \IEEEmembership{Senior~Member, IEEE}, Stylianos E. Trevlakis,~\IEEEmembership{Member, IEEE}, George C. Alexandropoulos, \IEEEmembership{Senior~Member, IEEE},\\ Theodoros A. Tsiftsis, \IEEEmembership{Senior~Member, IEEE}, and Rui Zhang, \IEEEmembership{Fellow, IEEE}
\thanks{E. Koutsonas and A.-A. A. Boulogeorgos are with the Department of Electrical and Computer Engineering, University of Western Macedonia, ZEP Area, 50100 Kozani, Greece (e-mail: \{dece00106, aboulogeorgos\}@uowm.gr).}
\thanks{S. E. Trevlakis is with the Department of Research and Development, InnoCube P.C., 55534 Thessaloniki, Greece. (e-mail: trevlakis@innocube.org)}
\thanks{G. C. Alexandropoulos is with the Department of Informatics and Telecommunications National and Kapodistrian University of Athens, Panepistimiopolis Ilissia, 16122 Athens, Greece (e-mail: alexandg@di.uoa.gr).}
\thanks{T. A. Tsiftsis is with the Department of Informatics and Telecommunications, University of Thessaly, Lamia 35100, Greece, and also with
the Department of Electrical and Electronic Engineering, University of Nottingham Ningbo China, Ningbo 315100, China (e-mail: tsiftsis@uth.gr)}
\thanks{R. Zhang is with 
the Department of Electrical and Computer Engineering, National University of Singapore, Singapore 117583 (e-mail: elezhang@nus.edu.sg).}
\thanks{Corresponding author: Alexandros-Apostolos A. Boulogeorgos (aboulogeorgos@uowm.gr).}
\thanks{The work of E. Koutsonas and A.-A. A. Boulogeorgos was supported by the research project MINOAS. The research project MINOAS is implemented in the framework of H.F.R.I call “Basic research Financing (Horizontal support of all Sciences)” under the National Recovery and Resilience Plan ``Greece 2.0'' funded by the European Union – NextGenerationEU (H.F.R.I. Project Number: 15857). The work of T. A. Tsiftsis and S. E. Trevlakis was supported by the project NEURONAS. The research project NEURONAS is implemented in the framework of H.F.R.I call ``3rd Call for H.F.R.I.’s Research Projects to Support Faculty Members \& Researchers'' (H.F.R.I. Project Number: 25726).}}

\maketitle
\vspace{-2cm}
\begin{abstract} 
In this paper, a new reconfigurable intelligent surface (RIS) hardware architecture, called self-organized RIS (SORIS), is proposed. The architecture incorporates a microcontroller connected to a single-antenna receiver operating at the same frequency as the RIS unit elements, operating either in transmission or reflection mode. The transmitting RIS elements enable the low latency estimation of both the incoming and outcoming channels at the microcontroller's side. In addition, a machine learning approach for estimating the incoming and outcoming channels involving the remaining RIS elements operating in reflection mode is devised. Specifically, by appropriately selecting a small number of elements in transmission mode, and based on the channel reciprocity principle, the respective channel coefficients are first estimated, which are then fed to a low-complexity neural network that, leveraging spatial channel correlation over RIS elements, returns predictions of the channel coefficients referring to the rest of elements. In this way, the SORIS microcontroller acquires channel state information, and accordingly reconfigures the panel's metamaterials to assist data communication between a transmitter and a receiver, without the need for separate connections with them. Moreover, the impact of channel estimation on the proposed solution, and a detailed complexity analysis for the used model, as well as a wiring density and control signaling analysis, is performed. The feasibility and efficacy of the proposed self-organized RIS design and operation are verified by Monte Carlo simulations, providing useful guidelines on the selection of the RIS elements for operating in transmission mode for initial channel estimation.       
\end{abstract}
    \vspace{-0.6cm}
\begin{IEEEkeywords}	
Reconfigurable intelligent surface (RIS), channel estimation, channel prediction, self-organization, spatial correlation. 
\end{IEEEkeywords}
    
	\vspace{-0.5cm}
\section{Introduction}\vspace{-0.2cm}
As we move towards the sixth-generation (6G) era of wireless networks, the need for supporting bandwidth-hungry applications, such as extended reality, metaverse, as well as the internet of senses, has steered the attention of the wireless communication community towards non-standardized higher frequency bands, i.e., millimeter wave, sub-terahertz (THz), THz, and optical~\cite{8387218,9583918,9615497}. However, high penetration losses in the aforementioned bands translate into blockages for communications~\cite{10914422}. Consequently, to mitigate their impact, advanced metamaterial-based structures named reconfigurable intelligent surfaces (RISs), or intelligent reflecting surfaces, which act as anomalous reflectors, have emerged~\cite{10596064,Pan_2022,10701743,EURASIP_RIS,10555049}. 

Building upon RISs, a generation of new system models has emerged that capitalizes on the concept of RIS real-time reconfiguration in order to create a favorable electromagnetic propagation environment. In this direction, extensive studies quantifying and optimizing the performance of RIS-empowered wireless systems have been pursued, e.g.,~\cite{8741198, Atapattu2020, 9367575, Koutsonas2023, 10930892, 9295369, 9437234, 10002889, 9629335, 10347404, 10263782, 10102423, 10049460, 10478511}. Among the first works that investigated the performance of RIS-empowered single- and multi-user wireless systems, under the assumption of perfect channel state information (CSI) knowledge, were~\cite{8741198, Atapattu2020}. Machine learning (ML) algorithms for RIS phase configuration, assuming perfect CSI, were devised in~\cite{9367575,Koutsonas2023,9864655,10930892}. In~\cite{9295369}, the uplink achievable rate was quantified for RIS-empowered millimeter-wave systems, assuming perfect CSI knowledge at the RIS. The authors of~\cite{9437234} introduced the simultaneous transmitting and reflecting (STAR) RIS concept, in which an RIS element can operate either in transmission or reflection mode. In~\cite{10002889}, the authors assumed statistical CSI at the RIS, for analyzing its performance and optimized the performance of an RIS-assisted eavesdropping wireless system. A weighted sum-rate maximization problem for STAR-RIS-assisted multiple-input multiple-output (MIMO) systems was formulated and solved in~\cite{9629335}. The authors of~\cite{10347404} assumed perfect CSI knowledge, and investigated the reliability and security of a non-orthogonal multiple access (NOMA) STAR-RIS system. Additionally, in~\cite{10263782}, a study on the outage probability, ergodic rate, and energy efficiency of the CR-STAR-RIS wireless system, assuming perfect CSI, was conducted. The authors of~\cite{10049460} presented a finite blocklength outage performance analysis for a STAR-RIS-aided downlink NOMA wireless system, assuming full and perfect CSI at the RIS controller. In~\cite{10478511}, the covert performance of a STAR-RIS wireless system employing rate-splitting multiple access was investigated, assuming perfect CSI availability at the RIS side. 

From the aforementioned contributions, it is evident that CSI knowledge is required at the RIS in order to make the most performance gains out of an RIS-empowered wireless system~\cite{9765815,Elsevier_CE}. To achieve this end, the authors of~\cite{8937491} proposed a channel estimation protocol for RIS-empowered orthogonal frequency division multiplexing (OFDM) systems, enabling the destination to estimate the cascaded source-RIS-destination channel, based on which the optimal phase shift of each RIS element was obtained. A channel estimation scheme for cyclic-prefixed single-carrier RIS-aided systems was presented in~\cite{10006744}. Channel estimation approaches for RIS-assisted multi-user wireless systems were designed in~\cite{9786794,9130088,9366805}. Again, the cascaded channel estimation was conducted at the destination, for which a link between the destination and the RIS microcontroller was required. In~\cite{9398559}, a two-stage cascaded channel estimation scheme was presented for RIS-aided millimeter wave multiple-input multiple-output wireless systems without a direct source-destination channel, using atomic norm minimization to sequentially estimate the channel parameters. After estimating the cascaded channel, at either the base station (BS) or the mobile station, a link is established between the estimator and the RIS microcontroller in order to feed the estimated channel to the latter. The authors of~\cite{9732214} presented a cascaded channel estimation strategy with low pilot overhead by leveraging the sparsity and the correlation of multiuser cascaded channels in millimeter-wave multiple-input multiple-output wireless systems, where the cascaded channel coefficients are estimated and then, sent to the RIS microcontroller. A low overhead RIS reconfiguration approach, which was based on designing the RIS phase shifts to illuminate the position where a user lies, thus, necessitating only user position estimation and not channel estimation, was proposed in~\cite{9673796}. Finally, the control overhead implications between RIS channel estimation schemes and beam management approaches~\cite{9129778,9827873} were analyzed in~\cite{10600711}. 

 {All the above works come with a common disadvantage: the need to inform the RIS microcontroller of the estimated channels by creating a dedicated feedback link, which poses challenges for practical implementation, such as eavesdropping~\cite{kunz2025lightweightsecurityambientpoweredprogrammable} and jamming~\cite{10143983}. Moreover, the communication latency between the open-reconfigurable intelligent controller of an open-radio access network and the RIS microcontroller may be 
microseconds, which is much higher than the typical  coherence time of mmWave wireless channels is in the order of microsecond. In other words, the channel information received by the RIS microcontroller may be outdated and 
no more relevant to the current CSI.} To avoid this, significant effort has been put into designing RISs with a small number of active components~\cite{9053976,9370097,10352433,10042447}. The authors in~\cite{9053976} were the first to propose waveguide-based connections of multiple RIS elements to reception radio frequency (RF) chains, designing a protocol, based on random RIS absorption configurations, for estimating the channel coefficients between each source and RIS element as well as between each RIS element and destination at the RIS's controller. This idea was later extended to hybrid RIS elements capable of simultaneous tunable reflection and absorption in~\cite{10352433}, where the received signal at the RIS side was used both for explicit~\cite{10042447} and parametric~\cite{9726785} channel estimation. This architecture enabled computationally autonomous and self-configuring metasurfaces~\cite{10352433}, and is recently considered together with energy harvesting for self sustainability~\cite{10693440}. In parallel, RISs comprising of mixed passive and active elements, mainly microsensors for the latter that were attached to reception RF chains, were proposed in~\cite{9370097} together with a compressed sensing and deep learning-based channel estimation scheme for predicting the CSI at RIS. In~\cite{9529045}, by leveraging the rank-deficient structure of RIS channels, the authors reported two practical residual neural networks, named single-scale enhanced deep residual and multi-scale enhanced deep residual, both capable of obtaining accurate CSI. In~\cite{9127834}, a deep denoising neural network assisted compressive channel estimation for millmeter-wave RIS-empowered wireless systems with reduced training overhead was introduced, which leverages the channel estimations extracted by a small number of active elements. The authors of~\cite{9511813} presented a  semi-passive elements-aided channel estimation framework for RIS, where a small portion of RIS elements were able to process the received signal for facilitating parametric channel estimation by employing estimation of signal parameters via super-resolution algorithms. In~\cite{10036135}, the authors presented  a channel estimation technique for RIS-aided multi-user multiple-input single-output wireless systems, in which the RIS is equipped with a small number of active elements. Finally, in~\cite{9882309}, a channel estimation technique with a single RF chain-based active RIS was proposed.  

 {
\begin{table*}
    \caption{SORIS novelty matrix}
    \centering
    \begin{tabular}{|c||p{1.5cm}|p{2cm}|p{2cm}|p{3cm}|c|}
        \hline
         Ref. &  RIS type & Number of RF chains & Energy consumption & Active elements selection &Functionality \\
         \hline \hline
         \cite{9053976} &  Hybrid & Multiple & Medium & Predetermined & Channel estimation\\
         \hline
         \cite{9370097} & Hybrid & Multiple & Medium & Predetermined & Sensing \\
         \hline
         \cite{10352433} & Hybrid & Multiple & Medium & Predetermined & Sensing \\
         \hline
         \cite{10042447} & Hybrid & Multiple & Medium & Predetermined & Channel estimation \\
         \hline
         \cite{9726785}  & Hybrid & Multiple & Medium & Predetermined & Localization \\
         \hline
         \cite{9529045} & Hybrid & Multiple & Medium & Predetermined & Channel estimation \\
         \hline
        \cite{9127834} & Hybrid & Multiple & Medium & Predetermined & Channel estimation \\
         \hline
         \cite{9511813} & Hybrid & Multiple & Medium & Predetermined & Channel estimation \\
         \hline
         \cite{10036135} & Hybrid & Multiple & Medium & Predetermined & Channel estimation \\
         \hline
         \cite{9882309}  & Hybrid & Multiple & Medium & Predetermined & Channel separation \\
         \hline
         SORIS & New & Single & Low & On-the-fly selection & Channel estimation \\
         \hline
    \end{tabular}
    \label{tab:table1}
\end{table*}}

 {The key disadvantage of RISs that are equipped with active elements is that each of these elements is connected to the microcontroller through a reception RF chain. Such a chain consists of a low-noise amplifier, a down-converter, and an analog-to-digital converter, which are all power-hungry units, as their average power consumption is usually in the order of several tens of milliwatt. As a result, as the number of active elements increases, so is the number of RF chains; hence, the power consumption increases towards unacceptable levels. Similarly holds for receiving RIS hardware architectures comprising only passive phase-tunable metamaterials~\cite{9053976,10352433}. Notably, an interesting work discusses an approach to acquire instantaneous CSI at a passive RIS without active elements is~\cite{9603291}, which is based on employing two anchor nodes that are deployed near the RIS. To the best of the authors' knowledge  {and as presented in Table~\ref{tab:table1}}, there is no existing work that presents an approach for acquiring instantaneous CSI at the RIS without significantly increasing the power consumption of the RIS, by employing either active elements at the RIS panel, active reception RF chains, or anchor nodes. Moreover, all the aforementioned works neglect the detrimental impact of hardware imperfections, which have been extensively discussed in~\cite{10535263,10844052,10502274}.} To cover the above gaps, this paper introduces a novel RIS hardware architecture in which the metasurface consists of elements capable of operating in both transmission and reflection modes, while its microcontroller is equipped by a receiver that operates at the same carrier frequency as the RIS elements. An efficient method with low complexity and by exploiting the dual-mode of the RIS elements is devised to enable CSI acquisition at the RIS controller. In more detail, a small number of RIS elements are selected, which are activated to operate in transmission mode in both the downlink and uplink. In what follows, we call the elements that operate in transmission mode as transmitting elements. The RIS controller is equipped by a single antenna and is able to capture the signals from the RIS transmitting elements. The controller first estimates the BS-RIS-controller channel and then divides it with the RIS-controller channel, which, due to the short distance of the RIS-controller link, is considered deterministic, thereby obtaining the required BS-RIS channels for the transmitting RIS elements. Following the similar process in the uplink, the user equipment (UE)-RIS channels at the selected transmitting elements are derived. Note that, based on the channel reciprocity principle, the UE-RIS and RIS-UE channels are assumed to be equal. Finally, by exploiting the spatial correlation among neighboring elements, we design a recurrent neural network (RNN) based algorithm for predicting the CSI for both the BS-RIS and RIS-UE links at the other unselected elements. To summarize, the main contributions of this paper are given as follows:
\begin{itemize}
    \item A novel RIS hardware architecture is proposed, which consists of elements that can operate in both transmit and reflection modes, as well as a microcontroller that is equipped with a single antenna receiver. The new architecture enables real-time channel estimation at the RIS microcontroller, thereby, achieving real-time self-organization and self-adaptation. We call this architecture self-organized reconfigurable intelligent surface (SORIS).   
    \item A new channel acquisition approach is designed, which leverages the transmit mode of a number of appropriately selected RIS unit elements to enable the estimation of their corresponding channels with both the BS and UE at the SORIS microcontroller side.
    \item A novel ML-empowered BS-RIS and RIS-UE channel estimation method is introduced, which takes the estimated channel coefficients for the selected transmit RIS elements by the microcontroller as input and generates an accurate estimation of the channel coefficients for the rest of the passive RIS elements as its output. 
    \item  {A complexity analysis for the used model, as well as a SORIS wiring density and control signaling analysis is performed, which highlights the applicability of the proposed architecture in realistic setups.}
    \item The relationship between the number and positions of the selected transmit RIS elements for initial channel estimation as well as the channel estimation accuracy of the proposed SORIS-based approach is investigated through Monte Carlo simulations. Insightful guidelines for selecting RIS elements to operate in transmission mode for the initial channel estimation are provided. 
    \item  {Finally, the impact of imperfect channel estimation due to the micro controller's receiver hardware impairments is investigated and assessed through Monte Carlo simulations.}
\end{itemize}

The rest of this paper is organized as: Section~\ref{S:SORIS} presents the system model, the SORIS architecture and its operation protocol. The channel estimation and prediction method is introduced in Section~\ref{S:Channel_Acquisition}, together with the performance metrics that are used to evaluate the efficiency of the proposed framework. Section~\ref{S:Results} presents our numerical results as well as key observations. Finally, the paper is concluded by summarizing the key findings in Section~\ref{S:Con}.  

\textit{Notations}: Small and capital bold letters are used to represent vectors and matrices, respectively. $\left|x\right|$ returns the absolute value of $x$, $x^p$ is the $p$-th power of $x$, and the $\rm{sinc}\left(x\right)$ function is given by $\frac{\sin(\pi\,x)}{\pi\,x}$ for $x\neq 0$, while $\rm{sinc}\left(0\right)=0$.  

\section{SORIS-Empowered System Model}\label{S:SORIS}

\begin{figure*}
    \centering
    \scalebox{0.60}{\hspace{-4cm}\input{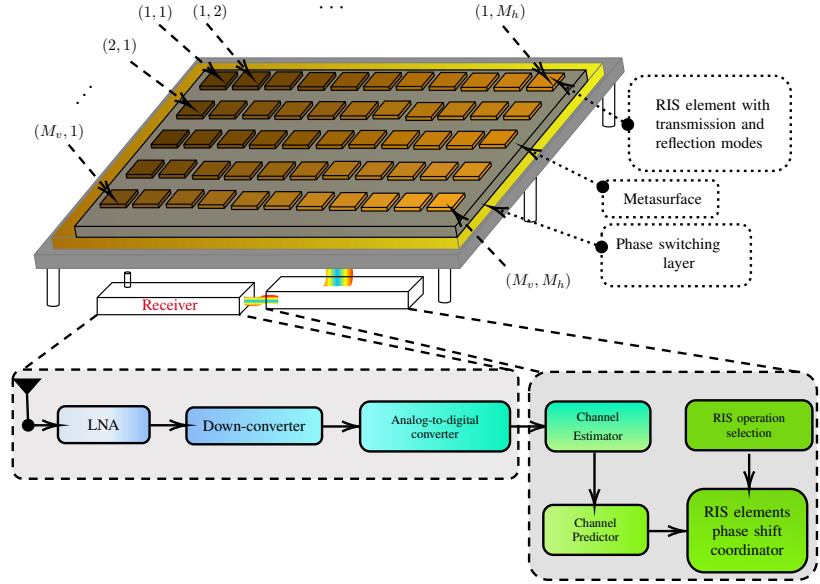}}
    \caption{The components of the proposed SORIS hardware architecture.}
    \hrulefill
    \label{Fig:SORIS}
\end{figure*}
\subsection{The Proposed SORIS Architecture}\label{SS:SORIS_arch}
The block diagram of the proposed SORIS hardware architecture is depicted in Fig.~\ref{Fig:SORIS}. Similar to conventional metasurfaces, the proposed surface architecture consists~of $N = M_vM_h$
metamaterials with reconfigurable responses. We define a grid-based coordinate system for the SORIS, according to which the first upper-left element is indexed by $(1, 1)$, the last upper-right element is indexed by $(1, M_h)$, while the element at the lowest-left part of the metasurface is indexed by $(M_v, 1)$, and that at the lowest-right part of the metasurface is indexed as $(M_v, M_h)$. Let the distance between neighboring elements, i.e., $(k,m)$ and $(k, m+1)$ as well as $(k,m)$ and $(k+1,m)$, be equal to $l$. Each element can be turned on/off independently of the rest metasurface elements. In particular, an element that is in ``on'' state can operate in either transmission or reflection mode. 

The operation mode of each SORIS element is imposed by a phase switching layer, which is connected through a wire to the SORIS microcontroller. This controller may be an application-specific integrated circuit (ASIC) low-power (LP) complementary metal-oxide semiconductor (CMOS)~\cite{9514889}, ASIC CMOS~\cite{Petrou2022}, ASIC fin field-effect transistor (FinFET) CMOS~\cite{8704523}, field programmable gate array (FPGA)~\cite{9551980}, computing processing unit (CPU)~\cite{Usman2022}, a graphical processing unit (GPU)~\cite{9690144}, or even a neuromorphic processing unit~\cite{10525226}. This controller can 
receive command to control the operation of the metasurface, such as  {beam steering}, beam spliting, and absorption, through an application programming interface. Moreover, it is responsible for the coordination of the on/off states of the unit elements, as well as their phase shifts when being under the ``on'' state.

In contrast to conventional RIS architectures~\cite{10596064,9053976,9370097,10352433,10042447,9603291}, in the SORIS microcontroller, a receiver that operates at the same band as the metasurface is equipped. The receiver is located behind the metasurface, with the task to capture signals relayed from the SORIS elements that have been switched to transmission mode. In particular, the captured signals are forwarded to the microcontroller, after passing through the receiver RF chain.  {Of note, the micro-controller is equipped with a single antenna. Thus, no mutual-coupling is considered at the receiver end. Notice that the assumption of a single antenna micro-controller is realistic due to the short distance between the active transmission elements and the receiver. Moreover, we assume that the microcontroller antenna and RF front-end are implemented with sufficient physical isolation, i.e. shielding through metallic enclosure, which allows minimization of the direct electromagnetic coupling and leakage from the near-field of the active RIS elements. This is a usual and realistic approach that is followed in most commercial receivers in order to guarantee  electromagnetic compatibility and mitigate the impact of power leakage. Finally, by assuming that the channel coefficient between the active RIS element and the microcontroller's antenna is directional and deterministic, i.e., it is known to the RIS microcontroller, digital signal processing (DSP) can be employed to fully mitigate the impact of hardware imperfections, including the deterministic part of leakage and coupling, as well as the effect of phase noise.} The microcontroller is responsible for estimating the corresponding baseband channel of the captured signal, as well as to predict the rest of the channels through leveraging the spatial correlation of high-frequency signals.   

In comparison with conventional RISs with multiple active reception elements~\cite{9053976,9370097,10352433}, each of which is connected to the microcontroller, SORIS employs only a single RF chain. As a consequence, its power consumption is significantly lower, and flexibility is increased.
In more detail, in the conventional RIS in~\cite{9370097} with a single RF chain per active element, the element that is used for channel estimation is predefined and fixed. On the contrary, in the proposed SORIS architecture, both the number of elements as well as the elements themselves can change dynamically in order to improve the channel acquisition accuracy. On the other hand, the approaches in~\cite{9053976,10352433} require in general more than one reception RF chain for explicit channel estimation~\cite{10042447}.

\vspace{-0.3cm}	
\subsection{System Model and Communication Protocol}\label{SS:SORIS_CaP}
\begin{figure*}
	\centering
    \scalebox{0.60}{\hspace{-4cm}\input{systemModel}}
	\caption{The considered SORIS-aided wireless communication system.}
	\label{Fig:SM}
   \hrulefill 
\end{figure*}
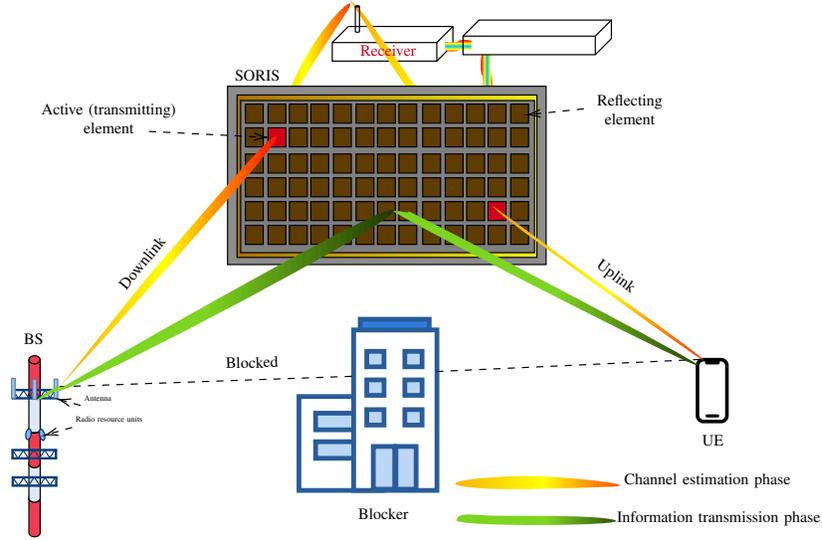

We consider a scenario in which the BS wants to establish a communication link with a UE through a SORIS, as depicted in Fig.~\ref{Fig:SM}. We assume that the BS is equipped with an analog beamformer that contains $N_t$ antenna elements, whereas the UE also realizes analog beamforming with $N_r$ antennas. The SORIS contains $N$ unit elements, which can be dynamically activated by its microcontroller. 

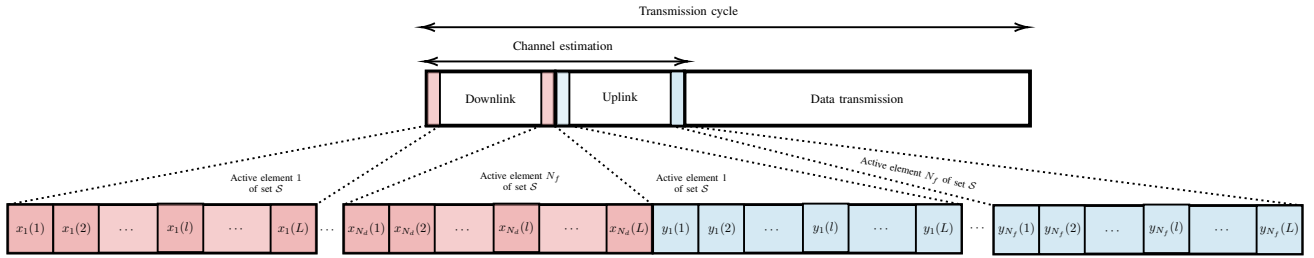
\begin{figure*}
	\centering
	\scalebox{0.48}{\input{commProt}}
	\caption{The communication frame of the proposed SORIS-empowered wireless system, including the estimation of the BS-SORIS (downlink) and SORIS-UE (uplink) channels.}
	\label{Fig:CommFrame}
    \hrulefill
\end{figure*}

The communication frame, illustrated in Fig.~\ref{Fig:CommFrame}, consists of two phases: one for channel estimation and the other for information transmission. It is assumed that the transmission cycle duration is lower than the channel coherence time. In other words, the BS-SORIS and SORIS-UE channel coefficients are almost static during a transmission~cycle. In the following, the proposed channel estimatio procedure is described.

\vspace{-0.3cm}
\section{Channel Acquisition with SORIS}\label{S:Channel_Acquisition}
\vspace{-0.3cm}
This section focuses on articulating the proposed channel acquisition approach with the SORIS architecture. In the following, the channel estimation protocol is presented in Section~\ref{SS:Channel_estimation_protocol}, while the ML-empowered channel prediction mechanism is described in Section~\ref{SS:Channel_prediction}.  

\subsection{Estimation of the Involved Channels}\label{SS:Channel_estimation_protocol}
The proposed channel estimation approach, which steps are summarized in Algorithm~\ref{Alg:Channel_coefficient_estimation}, begins by calculating the $N\times N$ correlation matrix, $\mathbf{C}$, between any two the channel coefficients $h_{k}$ and $h_{m}$ ($k,m=1, \ldots, N$), where $h_{k}$ is the channel coefficient between the BS and the $k$-th SORIS element, while $h_{m}$ stands for the channel coefficient between the BS and the $m$-th SORIS element. According to~\cite{Bjoernson2020}, the $(k,l)$-th element of the aforementioned correlation matrix can be obtained~as
\begin{align}
    {c}_{k,m}= \operatorname{sinc}\left(\frac{2\left\|{\mathbf{d}_{k,m}}\right\|}{\lambda}\right), 
    \label{eq::corellation}
\end{align}
where {$\mathbf{d}_{k,m}$ is the distance between} the $k$-th and $m$-th SORIS elements and $\lambda$ represents the signal wavelength. Notice that the correlation matrix only depends on the relative position of the RIS elements as well as the operation frequency, and not on the actual value of the BS-SORIS channel coefficients. Thus, the BS-SORIS correlation matrix of the BS-SORIS is equal to the correlation matrix of the SORIS-UE channels. 

Next, we select $N_f<N$ SORIS elements with the minimum absolute correlation values. For the sake of simplicity, we define the set of the selected elements as $\mathcal{S}=\{1,\ldots, N_f\}$.
In the downlink sub-phase of the channel estimation phase, we activate the SORIS element $s\in\mathcal{S}$, and set it in transmission mode, while the rest SORIS elements remain inactive. The BS transmits training signals towards the SORIS, which are captured by the SORIS microcontroller, after being relayed by the SORIS elements. The $l$-th ($l=1,\ldots,L$) baseband equivalent received sample at the microcontroller, when the SORIS elements is active and in transmission mode, can be expressed~as follows:
\begin{align}
    r_s(l) = h_{s} \, R_s \, g_s \, x_t(l) + n(l), 
    \label{eq:r_s}
\end{align}
where $h_s$ is the channel coefficient between the BS and the $s$-th SORIS element, which is modeled as {Rician} fading complex random variable with with factor $\kappa$, $R_s$ stands for the $s$-th SORIS element's response, which is known to the microcontroller, and $g_s$ is the $s$-th SORIS element to the microcontroller channel coefficient, which is also known to the RIS microcontroller. Moreover, $x_t(l)$ is the $l$-th BS beamformed transmitted signal and $n(l)$ represents the $l$-th sample of the additive white Gaussian noise, which is modeled as a zero-mean complex Gaussian process with variance $\sigma_n^2$. Recall that $L$ denotes the total number of training symbols that are transmitted by the BS, when the element $s\in\mathcal{S}$ is active.  {Of note, we assume that, as the channel between the RIS and the microcontroller is directional,  it can be considered  deterministic. As a result, the impact of hardware imperfections can be fully mitigated by employing DSP approaches. Finally, we assume that the microcontroller is equipped with a single antenna; thus, there is no mutual coupling at the microcontroller. On the other hand, the mutual coupling at the RIS has been already accounted for. } 


In the downlink sub-phase, the $L$ received signals at the microcontroller are organized in a matrix form defined~as
\begin{align}
    \mathbf{R} = \left[\begin{array}{c c c c}
         r_1(1) & r_1(2) & \cdots & r_{1}(L) \\
         r_2(1) & r_2 (2) & \cdots & r_{2}(L) \\
         \vdots & \vdots & \ddots & \vdots \\ 
         r_S(1) & r_S(s) & \cdots & r_{S}(L)
    \end{array}\right].
\end{align}
This matrix is fed to the microcontroller which employs a conventional channel estimation approach, e.g., minimum mean square error or least-square error, yielding the estimation
\begin{align}
   \tilde{f}_s = h_s \, R_s \, g_s.
   \label{eq:f_s}
\end{align}
Since $R_s$ and $g_s$ are known to the microcontroller, $h_s$ can be estimated~as follows:
\begin{align}
    \tilde{h}_s = \frac{\tilde{f}_s}{R_s\,g_s} 
    \label{Eq:tilde_h_s}
\end{align}
This process is repeated for all $s\in\mathcal{S}$, and its results are stored in the vector
\begin{align}
    \tilde{\mathbf{h}}= \left[\tilde{h}_1, \tilde{h}_2, \ldots, \tilde{h}_s, \ldots, \tilde{h}_S\right],
    \label{Eq:mathbf_h}
\end{align}
where $S$ is the size of $\mathcal{S}$. This is subsequently fed to the ML-empowered channel prediction engine, which is missioned to predict the overall BS-SORIS $N\times N_t$ channel matrix $\mathbf{H}$; we represent this prediction/estimation as $\tilde{\mathbf{H}}$. 


In the uplink sub-phase of the channel estimation phase, the UE steers its transmission beam towards the SORIS. Its microcontroller progressively activates one-by-one the SORIS elements that belong to the set $\mathcal{S}$, and sets their operation mode to transmission, while the rest of the elements remain inactive. The UE transmits beamformed training signals towards the SORIS, similarly to the BS, which pass through the activated metamaterials and are then captured by the microcontroller's receiver. The $l$-th sample ($l=1,\ldots,L_u$) of the baseband equivalent received signal at the microcontroller receiver, when the element $s\in\mathcal{S}$ is {activated}, can be expressed~as follows:
\begin{align}
    y_{s}(l) = h_{u,s}\,R_{u,s}\, g_s \, x_{u,t}(l) + n(l),
    \label{eq:y_s}
\end{align}
where $h_{u,s}$, $R_{u,s}$, and $g_{s}$ respectively stand for the UE-$s$-th SORIS element channel coefficient, the $s$-th SORIS element response, and the channel coefficient between the $s$-th SORIS element and microcontroller, while $x_{u,t}(l)$ is the $l$-th transmitted signal by the UE. Note that $L_u$ represents the total number of training symbols that are transmitted by the UE, when the element $s\in\mathcal{S}$ is {activated}.  

In the uplink sub-phase, the received signals at the microcontroller are organized in the following matrix: 
\begin{align}
    \mathbf{Y} = \left[\begin{array}{c c c c}
         y_1(1) & y_1(2) & \cdots & y_{1}(L) \\
         y_2(1) & y_2 (2) & \cdots & y_{2}(L) \\
         \vdots & \vdots & \ddots & \vdots \\ 
         y_S(1) & y_S(s) & \cdots & y_{S}(L)
    \end{array}\right].
\end{align}
Again, the microcontroller implements the same channel estimation technique and returns an estimation of the UE-SORIS-microcontroller end-to-end channel, which is expressed~as
\begin{align}
   \tilde{f}_{u,s} = h_{u,s}\,R_{u,s}\, g_s
   \label{Eq:f_u_s}
\end{align}
for each  UE-$s$-th SORIS element channel coefficient.
Then, the estimation of the UE-$s$-th SORIS element is obtained~as follows:
\begin{align}
    \tilde{h}_{u,s} = \frac{\tilde{f}_{u,s}}{R_{u,s}\, g_s}.
    \label{Eq:tilde_h_u_s}
\end{align}
This process is repeated for all $s\in\mathcal{S}$ active elements of the SORIS. Note that based on the reciprocity principle~\cite{Gao2024,Xu2024}, the UE-$s$-th SORIS element channel coefficient is equal to the  $s$-th SORIS element-UE channel coefficient. Finally, the channel estimations of the UE-$s$-th SORIS element channel coefficients are stored into the vectors 
\begin{align}
    \tilde{\mathbf{h}}_{u} = \left[ \tilde{h}_{u,1}, \tilde{h}_{u,2}, \ldots, \tilde{h}_{u,s}, \ldots, \tilde{h}_{u,S}  \right].
    \label{Eq:tilde_mathbf_h_u}
\end{align}
As before, $\tilde{\mathbf{h}}_u$ is fed to the ML-empowered channel prediction engine, which returns a prediction of the SORIS-UE $N_r\times N$ channel matrix $\mathbf{H}_u$; we represent this prediction as $\tilde{\mathbf{H}}_u$.

\begin{algorithm}
\caption{SORIS Channel Acquisition and Prediction}
\label{Alg:Channel_coefficient_estimation}
\algsetup{linenosize=\small}
\small
\begin{algorithmic}[1]
\renewcommand{\algorithmicrequire}{\textbf{Input:}}
\renewcommand{\algorithmicensure}{\textbf{Output:}}
\REQUIRE RIS element positions, carrier frequency $\lambda$, Set of active transmission-mode elements $\mathcal{S}$ (size $N_f$), Training symbol count $L$, $L_u$, Known responses $R_s$, $R_{u,s}$, $g_s$ $\forall s \in \mathcal{S}$, Pre-trained RNN model
\ENSURE Estimated channel matrices: $\mathbf{\tilde{H}}$, $\mathbf{\tilde{H}}_u$
    
    \medskip
    \textit{Initialization:}
    \STATE Compute spatial correlation matrix $\mathbf{C} \in \mathbb{C}^{N \times N}$:
    \FOR{$k = 1$ \TO $N$}
        \FOR{$m = 1$ \TO $N$}
            \STATE $c_{k,m} \gets$ Eq.~\eqref{eq::corellation}
        \ENDFOR
    \ENDFOR
    
    \medskip
    \textit{Channel Estimation:}
    \STATE $\mathbf{\tilde{h}} \gets \text{empty vector}$, $\mathbf{\tilde{h}}_u \gets \text{empty vector}$
    
    \textit{Downlink (BS-RIS):}
    \FOR{each $s \in \mathcal{S}$}
        \STATE Set element $s$ to reception mode
        \STATE Transmit symbols $\mathbf{x}_t = [x_t(1),\dots,x_t(L)]$
        \STATE $r_s(l) \gets$ Eq.~\eqref{eq:r_s}
        \STATE $\tilde{f}_s \gets$ Eq.~\eqref{eq:f_s}
        \STATE $\tilde{h}_s \gets$ Eq.~\eqref{Eq:tilde_h_s}
        \STATE Append $\tilde{h}_s$ to $\mathbf{\tilde{h}}$
    \ENDFOR
    
    \textit{Uplink (UE-RIS):}
    \FOR{each $s \in \mathcal{S}$}
        \STATE Set element $s$ to transmission mode
        \STATE Transmit symbols $\mathbf{x}_{u,t} = [x_{u,t}(1),\dots,x_{u,t}(L_u)]$
        \STATE $y_s(l) \gets$ Eq.~\eqref{eq:y_s}
        \STATE $\tilde{f}_{u,s} \gets$ Eq.~\eqref{Eq:f_u_s}
        \STATE $\tilde{h}_{u,s} \gets$ Eq.~\eqref{Eq:tilde_h_u_s}
        \STATE Append $\tilde{h}_{u,s}$ to $\mathbf{\tilde{h}}_u$
    \ENDFOR
    \medskip
    \textit{Channel Prediction:} \\
    \STATE $\mathbf{a} \gets$ Eq.~\eqref{Eq:a_vec}
    \STATE $\mathbf{a_u} \gets$ Eq.~\eqref{Eq:a_vec}
    \STATE Apply spatial stamps $z_i$ (element positions)
    \STATE $\mathbf{\tilde{H}} \gets \text{RNN}(\mathbf{a})$
    \STATE $\mathbf{\tilde{H}_u} \gets \text{RNN}(\mathbf{a_u})$
    
    \medskip
    \RETURN $\mathbf{\tilde{H}}$, $\mathbf{\tilde{H}}_u$
\end{algorithmic}
\end{algorithm}
\subsection{Channel Extrapolation with ML}\label{SS:Channel_prediction}

The objective of the neural network in the proposed SORIS architecture is to provide CSI predictions for the BS-$i$-th SORIS element and the $i$-th SORIS element-UE, where $i\notin\mathcal{S}$ indicates the index of the inactive metamaterial. In neural networks, the size of the training dataset is proportional to the number of parameters to be trained~\cite{8949757}. Motivated by this, we employ a low-complexity recurrent neural network (RNN), which is capable of processing sequential data by detecting patterns in the sequences. 

\begin{figure*}
	\centering
    \scalebox{0.4}{\input{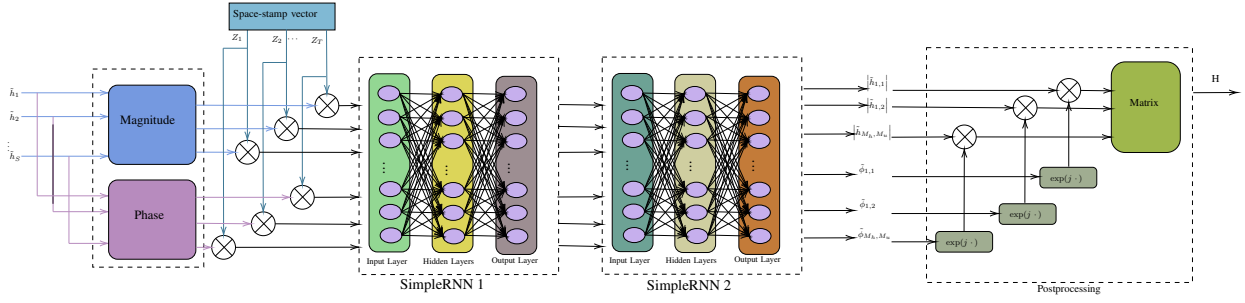}}
	\caption{The designed ML-empowered channel prediction engine for our SORIS architecture.}
	\label{Fig:ML}
    \hrulefill 
\end{figure*}

A graphical representation of the network structure that we use is presented in Fig.~\ref{Fig:ML}. The network consists of three units: \textit{i}) preprocessing, \textit{ii}) RNN, and \textit{iii}) postprocessing. The preprocessing unit receives as input the elements of $\tilde{\mathbf{h}}$ and outputs the following augmented vector:
\begin{align}
    \mathbf{a} = \left[ \left|\tilde{h}_1\right|, \left|\tilde{h}_2\right|, \ldots, \left|\tilde{h}_S\right|, \tilde{\theta}_1, \tilde{\theta}_2, \ldots, \tilde{\theta}_S  \right],
    \label{Eq:a_vec}
\end{align}
where $\tilde{\theta}_i$, with $i=1,\ldots, S$, is the phase of $\tilde{h}_i$. Next, a space-stamp, $z_i$ with $i\in[1, S]$, for each of the elements of \eqref{Eq:a_vec} is multiplied with the corresponding value of $\mathbf{a}$. The space-stamp is a matrix that describes the position of each element of the set $\mathcal{S}$ in the SORIS grid. Notice that $\left|\tilde{h}_i\right|$ and $\tilde{\theta}_i$ have the same space-stamp $z_i$. 


The preprocsessing unit forwards its $2S$ outputs to the RNN. A many-to-many without synced sequence input RNN network is assumed. 
The network consists of three layers, a recurrent and two feed-forward layers. As a recurrent layer, we use a SimpleRNN layer. Subsequently, two dense layers with $R_{d,1}$ and $R_{d,2}$ units forward connected are used as feed-forward layers. The SimpleRNN network consists of an input layer with $N$ units, $1$ hidden layers with $R_h$ units, and an output layer with $R_u$ units. In terms of the compiling methods, a stochastic gradient descent optimizer with a learning rate of $L_r$ is~used. 

\subsection{Complexity analysis}

This section focuses on analyzing the complexity of Algorithm~\ref{Alg:Channel_coefficient_estimation}. Algorithm~\ref{Alg:Channel_coefficient_estimation} consists of three phases, i.e. i) initialization phase, ii) channel estimation phase and iii) channel prediction phase. In the initialization phase, two nested loops; thus, the complexity of this phase is equal to $\mathcal{O}(N^2)$. The channel estimation phase consists of two sub-phases: i) downlink (BS-RIS) channel estimation and ii) uplink (UE-RIS) channel estimation. The complexity of the downlink channel estimation sub-phase is equal to $\mathcal{O}\left(N_f\times L\right)$, while the complexity of the uplink channel estimation sub-phase is $\mathcal{O}\left(N_f\times L_u\right)$. As a result, the complexity of the channel estimation phase is equal to $\mathcal{O}\left(N_f\times \left(L+L_u\right)\right)$ In the channel prediction phase, each RNN employs $1$ hidden layer and process a sequence of length $N_f$. As a consequence, the complexity of the channel prediction phase is $\mathcal{O}\left(N_f\right)$. Building upon the above observations, the overall complexity of Algorithm~\ref{Alg:Channel_coefficient_estimation} can be expressed as $\mathcal{O}\left(N^2+N_f\left(L+L_u+1\right)\right)$. 

 {The training complexity of the recurrent neural network (RNN) can be evaluated as 
        $\mathcal{O}_t = \mathcal{O}\left(N_e\,N_s\,N\,L_H^2\right),$
    where $N_e$, $N_s$ are respectively the number of epochs and the total number of channel samples. Notice that for Algorithm 1, the training complexity is incurred only once, offline, before the main algorithm is executed. The final step of the provided algorithm is simply the inference (prediction) step, which has a much lower complexity of $\mathcal{O}(N\times R_h)$ per execution.}

     {The space complexity is determined by the data structures that are stored. In more detail, we store the spatial correlation matrix, of which the size is proportional to $N^2$, the input and output channel vectors of size $N$, and the pre-trained RNN model weights, with size proportional to $N\times R_h + R_h^2$. The temporary storage that is required between the loops is usually much smaller that the channel vector size; thus, it can be neglected. The overall memory footprint of Algorithm 1 is thus determined by the storage of the dominant term which is either the spatial correlation matrix or the size of the pre-trained RNN model, and it can be evaluated as $\mathcal{O}\left(\max(N^2, N\,R_h+R_h^2)\right)$.}

\subsection{Wiring density and control signaling analysis}\label{SS:Performance_metrics}

 {The total number of wires can be obtained as $W_t = N\, B_p + N_f \, B_m,$
          where  $B_p$ stands for the phase - shift bits, and $B_m$ is the mode bits.  The control latency can be evaluated as $T_c = \max(T_s, T_w),$ where $T_s$ stands for the signaling overhead, and $T_w$  is the switching latency. Of note, $T_s << T_w$; thus, $T_c = T_s.$ The signaling overhead can be obtained as
              $T_s = \frac{W_t}{R_b},$  where $R_b$ is the digital bus data rate.}

           {\textit{Example:} In the case of $N=256$ elements, by assuming $B_p=2$, $B_m=1$, and $N_f=8$, we have $W_t=520$ wires, while in the case of $N=1024$ we have $W_t=2052$ wires.  By assuming that a differential signaling approach is employed, like double date rate interface, or low-voltage differential signaling (LVDS), $R_b$ is in the order of $1\,\mathrm{Gbps}$. Thus, for $N=256$ RIS elements, $T_s=0.52\,\,\mu s$, while for $N=1024$, $T_s=2.052\,\,{\mu s}$, which are both much lower than the coherence time of typical mmWave channels.}
\subsection{Performance Metrics}\label{SS:Performance_metrics}
To quantify the performance of the proposed SORIS communication protocol, we consider two metrics: \textit{i}) CSI acquisition latency; and \textit{ii}) average mean square error (AMSE). We assume that the transmission period of a single training symbol is equal to $T_s$; since the number of samples that are needed in order to estimate is $L$ and the number of transmit elements that are used at the SORIS is $N_f$. The CSI acquisition latency can be obtained as
\begin{align}
    D = 2 \, N_f \, L \, T_s. 
    \label{Eq:D}
\end{align}
From~\eqref{Eq:D}, we observe that the CSI acquisition latency depends on the efficiency of the channel estimator, i.e., the number of received signal samples that it requires for an accurate estimation ($N_f$), the number of transmit elements that are employed, and the symbol duration. 

In a single channel prediction, the mean square error (MSE) of the channel coefficients magnitude is defined~as follows:
\begin{align}
    e_{h} = \frac{1}{N}\sum_{n_h=1}^{M_h}\sum_{n_v=1}^{M_v}\left( \left|h_{n_h,n_v}^{\chi}\right|-\left|\tilde{h}_{n_h,n_v}^{\chi}\right|\right)^2,
    \label{Eq:e_h}
\end{align}
where $\tilde{h}_{n_h,n_v}^{\chi}$ is the $(n_h,n_v)$-th element of the prediction matrix $\tilde{\mathbf{H}}_u^{\chi}$, while ${h}_{n_h,n_v}^{\chi}$ is the actual channel coefficient. Moreover, ${\chi}=\rm{d}$, if we refer to the BS-SORIS link, or ${\chi}=\rm{u}$, if we refer to the SORIS-UE link. 
Similarly, the MSE of the channel coefficients phase is defined~as
\begin{align}
    e_{\theta} = \frac{1}{N}\sum_{n_h=1}^{M_h}\sum_{n_v=1}^{M_v}\left( \left|\theta_{n_h,n_v}^{\chi}\right|-\left|\tilde{\theta}_{n_h,n_v}^{\chi}\right|\right)^2,
    \label{Eq:e_theta}
\end{align}
where $\tilde{\theta}_{n_h,n_v}^{\chi}$ is the $(n_h,n_v)$-th element of the phase of the prediction matrix $\tilde{\mathbf{H}}_u^{\chi}$, while ${\theta}_{n_h,n_v}^{\chi}$ is the actual phase of the corresponding channel coefficient. The AMSE of the channel coefficients magnitude and phases can be obtained by averaging~\eqref{Eq:e_h} and~\eqref{Eq:e_theta}, respectively.  

\section{Numerical Results and Discussion}\label{S:Results}
In this section, we present numerical results for the channel estimation performance of the proposed SORIS framework.
The results are accompanied with insights on the the trade-off between accuracy and channel acquisition latency, leading to practical guidelines for the maximization of the channel predictor's performance through selecting the most appropriate SORIS elements for being in the active transmission mode. To this end, Section~\ref{SS:results} includes our {Monte}-Carlo-based AMSE results for channel estimation, while Section~\ref{SS:Guidelines} elaborates on the selection of the active SORIS elements.

\subsection{Channel Estimation Performance}\label{SS:results}
We first consider a SORIS with $8\times8$ elements. All the {Rician} channel coefficients are generated with $\kappa=8$~dB. At the RNN network training phase, a learning rate of a $L_r=0.001$ is used and $E_p=100$ epochs were simulated. The training batch size is set to $B_s=32$. Four simulation scenarios are considered. In the first scenario, the $4$ {edge} elements of the RIS are assumed active.  In the second, $4$ {edge} elements and $4$ central elements of the SORIS are considered as active.  At the third simulation scenario, a grid of $16$ transmit elements with $48$ non-transmit ones is used. The same grid pattern as the third scenario is followed in the fourth scenario, with $32$ transmit and $32$ non-transmit elements.

Fig.~5 illustrates the AMSE of the channel coefficient magnitude  as a function of number of transmit elements for different correlation matrices. From this figure, we observe that for a fixed number of active elements, as the correlation between neighbor elements increases, i.e. the inter-element distance decreases, the  AMSE decreases. Additionally, for a given inter-element distance, as the number of active elements increases, the  AMSE decreases. 

\begin{figure}
\centering
    \begin{minipage}{.45\textwidth}
	\centering
	 {\scalebox{0.75}{\input{Abs}}
	\caption{AMSE of the channel magnitude at different SORIS transmit elements.}}
	\label{Fig:fig2}
    \end{minipage}
    \hspace{+0.2cm}
    \begin{minipage}{.45\textwidth}
    \centering
	\scalebox{0.75}{\input{Angles}}
	\caption{AMSE of the channel phase at different SORIS transmit elements.}
	\label{Fig:fig3}
    \end{minipage}
\end{figure}

Fig.~6 illustrates the  AMSE of the channel coeffiecient phase at different correlation matrices as a function of number of transmit elements. It becomes evident that for a given number of active elements, as the correlation between neighbor elements increases, i.e. the inter-element distance decreases, the  AMSE also decreases. Finally, for a given inter-element distance, as the number of active elements increases, the  AMSE decreases.

\begin{figure}
    \centering
	\scalebox{0.85}{\input{16x16n32x32Angles}}
	\caption{AMSE of the channel magnitude and phase vs the number of SORIS transmit elements, for different values of $N$.}
	\label{Fig:fig1024}
\end{figure}
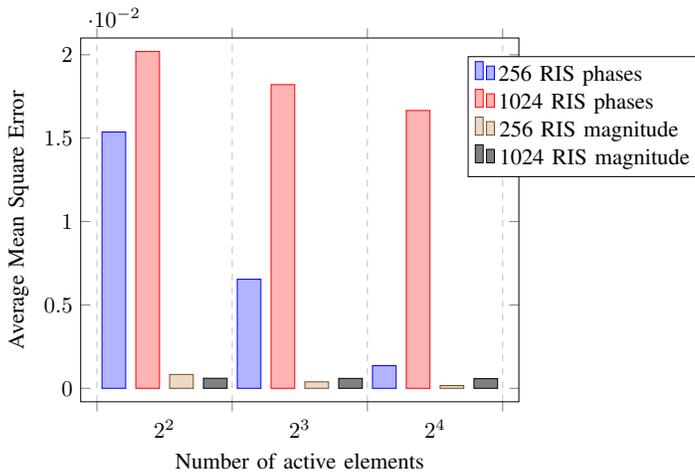
 {In Fig.~\ref{Fig:fig1024}, the AMSE of the channel magnitude and phase as a function of the number of SORIS transmit elements, for different values of $N$ is illustrated. From this figure, we verify again that, for a given $N$, as the number of active elements increases,   both the AMSE of the channel magnitude and the AMSE of the phase decrease. Moreover, for a given number of active elements, as $N$ increases, both the AMSE of the channel magnitude and the AMSE of the phase also increase. This indicates that in order to improve the accuracy of the model, the number of active elements should scale in the same manner as the size of the SORIS.}


\begin{figure}
	\centering
    \subfloat[]{\scalebox{0.3}{\input{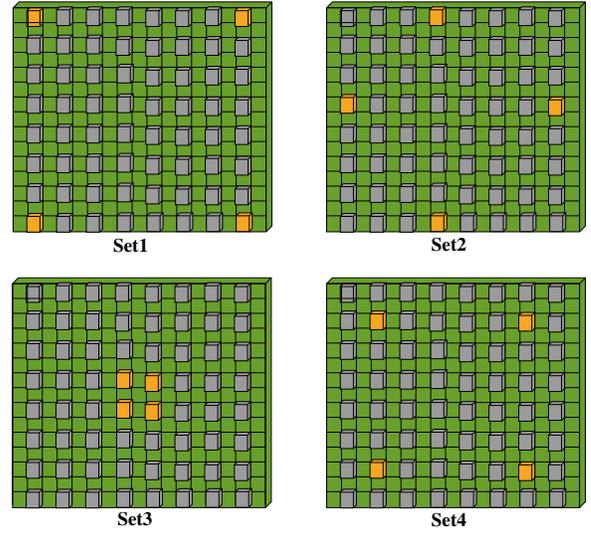}
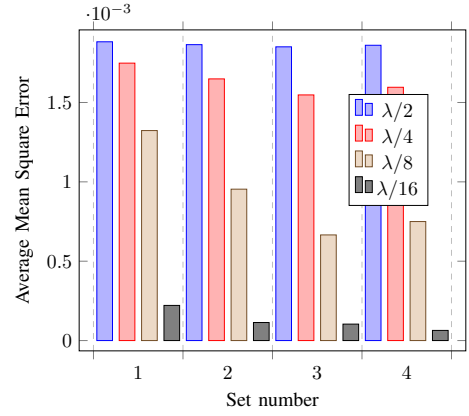
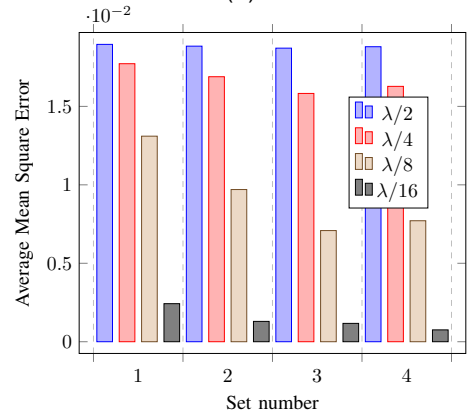}\label{fig:4active_sets}}\\
    \subfloat[]{\scalebox{0.75}{\input{4active_Abs}}\label{Fig:fig4}} \hspace{+0.1cm}
    \subfloat[]{\scalebox{0.75}{\input{4active_Angle}}\label{Fig:fig5}}
	\caption{(a) Sets of SORIS transmit elements, (b) AMSE of the channel coefficients' magnitude, and (c) AMSE of the phases of the the channel coefficients.}
\end{figure}

In the previous estimation results, the SORIS active transmit elements were selected in a random fashion, i.e. without performing any optimization approach. However, it is intuitively expected that this selection, when optimized, can result in AMSE performance improvement. Motivated by this, we consider the scenario in which only $4$ elements are active, and we define $4$ indicative sets. A graphical representation of the sets is presented in Fig.~\ref{fig:4active_sets}.  In this figure, orange and gray colors are used for the transmit and non-transmit elements, respectively. For example, in the first set, the transmit elements are $(1,1), (1,8), (8,1)$ and $(8,8)$. This scenario corresponds to the case in which the maximum possible distance between the active elements is achieved. As a result, the minimum average correlation between the active units is expected. From Fig.~\ref{Fig:fig4}, for $l=\lambda/2$, we observe that the AMSE of the channel coefficient magnitude prediction is in the order of $6\times 10^{-2}$, while the corresponding AMSE of the channel coefficient angle prediction is in the order of $6\times 10^{-2}$, as depicted in Fig.~\ref{Fig:fig5}. Moreover, it is shown for set $1$ that, as the distance between neighbor elements decreases, their correlation increases; thus, the accuracy of the proposed ML prediction engine increases. In the second set, the transmit elements are the following: $(1,4)$, $(4,1)$, $(4,8)$, and $(8,4)$. Let us consider the scenario in which the distances between neighboring elements is equal to $\lambda/2$. The correlation between  elements $(1,4)$ and $(4,1)$ is $0.05$, while the correlation between $(1,4)$ and $(4,8)$ is approximately $0.07$. The correlation between $(4,8)$ and $(8,4)$ is equal to $0.05$, whereas the correlation between $(8,4)$ and $(4,1)$ is approximately $0.05$. The correlations between $(1,4)$ and $(8,4)$ as well as $(4,1)$ and $(4,8)$ are both  $0$. The correlation between each transmit element and the closest neighboring elements is equal to $0$. Since there is no important change in the correlation of the elements in comparison to the first set, for $\lambda/2$, we expect similar values for both the AMSE for the magnitude and phase of the channel coefficients. This intuition is verified in Figs.~\ref{Fig:fig4} and~\ref{Fig:fig5}. However, as the distance between two neighboring elements decreases, the correlations increase, leading to an increase on the prediction accuracy. From Fig.~\ref{Fig:fig4}, we observe that the AMSE of the channel coefficient magnitude decreases from $4.37\times 10^{-3}$ to $3.18\times 10^{-3}$ by selecting set $2$ instead of set $1$, for closest neighboring element distance equal to $\lambda/8$. Similarly, for the same distance between the closest neighboring elements, it becomes evident from Fig.~\ref{Fig:fig5} that the AMSE of the channel coefficient phase decreases from $4.32\times 10^{2}$ to $3.23\times 10^{-2}$ by selecting set $2$ instead of set~$1$. The third set has the following transmit elements $(4,4)$, $(4,5)$, $(5,4)$, and $(5,5)$. According to Fig.~\ref{Fig:fig4}, for $l=\lambda/8$, in terms of the AMSE of the channel coefficient amplitude, set $3$ outperforms sets $1$ and $2$ by $46.5\%$ and $25.8\%$ AMSE reduction, respectively, while, based on Fig.~\ref{Fig:fig5}, in terms of the AMSE of the channel coefficient phase, set $3$ outperforms sets $1$ and $2$ by $44.19\%$ and $25\%$ AMSE reduction, respectively. Finally, let us investigate the case in which the distance between the closest neighbors is equal to $\lambda/16$. In this case, the correlation between the first and second order neighboring elements increases and is higher of $0.95$. This is translated in a slight AMSE increase, as shown in Figs.~\ref{Fig:fig4} and~\ref{Fig:fig5}. In the fourth set, the transmit elements are the following: $(2,2)$, $(2,7)$, $(7,2)$, and $(7,7)$. For the case of $l=\lambda/4$, the correlation between $(2,2)$ and $(2,7)$ as well as $(7,2)$ equals $0.12$, while the correlation between $(2,2)$ and $(7,7)$ is $0.09$. Additionally, the correlation between $(2,7)$ and $(7,7)$ is $0.12$, while the correlation between $(2,7)$ and $(7,2)$ is $0.09$. The correlation between $(7,2)$ and $(7,7)$ is equal to $0.12$. The correlations between the transmit elements and the first and second order neighboring elements are equal to $0.66$ and $0.36$, respectively. Notice that each transmit element has $4$ first-order neighboring elements and $4$ second-order neighboring elements. Despite the fact that, in this setup, the transmit elements have the maximum number of neighboring elements, no signification performance increase in comparison to the third set is observed, due to the relatively low correlation between neighboring elements. Despite the low correlation, the achievable AMSE for the channel coefficients magnitude and phase are acceptable. 

\begin{figure}
	\centering
    \subfloat[]{\scalebox{0.3}{\input{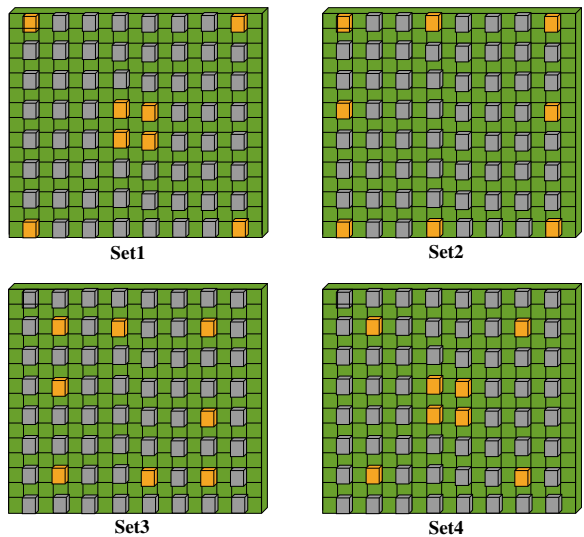}
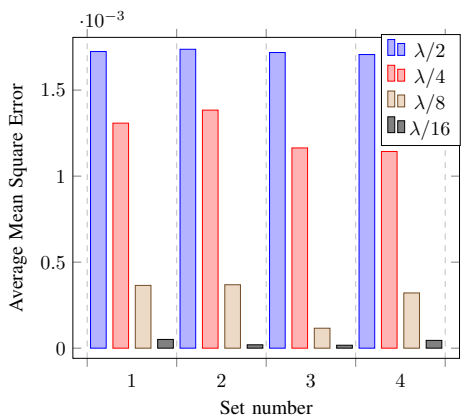
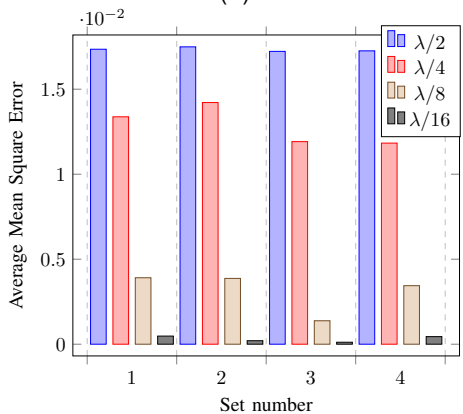}\label{fig:8active_sets}}\\
    \subfloat[]{\scalebox{0.75}{\input{8active_Abs_new1}}\label{Fig:fig6}}\hspace{+0.1cm}
    \subfloat[]{\scalebox{0.75}{\input{8active_Angle_new1}}\label{Fig:fig7}}
	\caption{(a) Sets of transmit elements, (b) AMSE of the channel coefficients' magnitude, and (c) AMSE of phases of the channel coefficients.}
    \label{Fig:8elements}
\end{figure}

A scenario in which $8$ SORIS elements are subsequently activated is presented in Fig.~\ref{Fig:8elements}. We have particularly considered four different sets. In the first set, the following SORIS elements are activated: $(1,1)$, $(1,8)$, $(4,4)$, $(4,5)$, $(5,4)$, $(5,5)$, $(8,1)$, and $(8,8)$. Element $(1,1)$ has $2$ first-order neighboring elements, i.e., $(1,2)$ and $(2,1)$, and $1$ second-order neighboring element, i.e., $(2,2)$. Similarly, each of the $(1,8)$, $(8,1)$, and $(8,8)$ elements has $2$ first-order neighboring elements and $1$ second-order neighboring element. Element $(4,4)$ has $2$ passive first-order neighboring elements, i.e., $(4,3)$ and $(3,4)$ and $3$ second-order neighboring elements, i.e., $(3,3)$, $(3,5)$, and $(5,3)$. Additionally, each of the $(4,5)$, $(5,4)$, and $(5,5)$ elements have $2$ passive first-order neighboring elements and $3$ second-order neighboring elements. We have investigated four scenarios: the first scenario with $l=\lambda/2$, the second with $l=\lambda/4$, while $l=\lambda/8$ and $l=\lambda/16$ for the third and fourth scenarios, respectively. in the first scenario, by comparing the correlations between the first-order neighboring elements of the transmit elements of the aforementioned sets, we observe that, if the distance between neighboring elements is $\lambda/2$, the correlation is $0$ in all the aforementioned sets. However, we observe a slight AMSE degradation when employing the first set of Fig.~\ref{fig:8active_sets} instead of either the first or the third set of Fig.~\ref{fig:4active_sets}. This is due to the fact that the number of the second-order neighboring elements to transmit elements is higher in the first set of Fig.~\ref{fig:8active_sets}, compared to the first and the third sets of Fig.~\ref{fig:4active_sets}. Next, let us focus in the case where the distance between neighboring elements is equal to $\lambda/4$. Excluding the correlations between $(4,4)$, $(4,5)$, $(5,4)$, and $(5,5)$ elements, we observe that the correlations between the rest transmit elements are relatively small. As a result, its transmit element provides information for a different part of the SORIS. Moreover, the correlation of the transmit elements with the first order neighboring elements is relatively high, i.e., higher than $0.5$; hence, the accuracy of the prediction engine is expected to increase. Indeed, from Fig.~\ref{Fig:fig6}, we observe that, as the distance between neighboring elements decreases, both AMSEs of the magnitude and phase decrease. For the case of $l=\lambda/8$, the average correlation between the SORIS passive and transmit elements increases. As a result, the prediction accuracy of the ML engine increases; thus, AMSEs for both the magnitude and phase decrease. In the case of $l=\lambda/16$, we observe that both the correlation between neighboring elements and the correlation between the transmit elements increase, as the distance of the neighboring elements decreases. As a consequence, the performance of the channel prediction engine improves. In more detail, as depicted in Fig.~\ref{Fig:fig6}, by comparing the case the neighboring distance is $\lambda/8$ with the one in which the neighboring distance is $\lambda/16$, the AMSE of the magnitude decreases from $3.65\times 10^{-4}$ to $5.08\times 10^{-5}$, i.e., a more than one order of magnitude decrease is observed. Similarly, from Fig.~\ref{Fig:fig7}, we see that the AMSE of the magnitude decreases from $3.91\times 10^{-3}$ to $4.79\times 10^{-4}$, as the distance of the neighboring elements decreases from $\lambda/8$ to $\lambda/16$.      

The results with second set having the following transmit elements: $(1,1)$, $(1,4)$, $(1,8)$, $(4,1)$, $(4,8)$, $(8,1)$, $(8,4)$, and $(8,8)$, are presented in~Fig.~\ref{fig:8active_sets}. Each of the $(1,1)$, $(1,8)$, $(8,1)$, and $(8,8)$ elements has $2$ first-order and $1$ second-order neighboring elements. Each of $(1,4)$, $(4,1)$, $(4,8)$, and $(8,4)$ has $3$ first-order and $2$ second-order neighboring elements. For $l=\lambda/2$, we observe that the average correlation between the active and passive elements decreases in comparison to the first set; as a result, the AMSEs for both the channel coefficient magnitude and phase increase. The same observation can be made for $l=\lambda/4$. On the other hand, for $l=\lambda/8$, it is shown that the average correlation between the active and passive elements increases in comparison to the corresponding scenario of the first set; as a consequence, the AMSEs for both the channel coefficient magnitude and phase decrease. Finally, for  $l=\lambda/16$, the average correlation between the active and passive elements increases in comparison to the corresponding scenario of the first set; thus, the AMSEs for both the channel coefficient magnitude and phase decrease.    

In the third set, the following elements are active: $(2,2)$, $(2,4)$, $(2,7)$, $(4,2)$, $(5,7)$, $(7,2)$, $(7,5)$, and $(7,7)$. Each transmit element has $4$ first-order neighbors and $4$ second-order neighboring elements. We observe that the average correlation between the transmit elements of set $3$ is lower than the average correlation between the transmit elements of both sets $1$ and $2$. As a consequence, each transmit element provides information of a different part of the SORIS; hence, the prediction accuracy is expected to increase. This is verified by the results of Figs.~\ref{Fig:fig6} and~\ref{Fig:fig7}. Similar results are observed for the case where the distance between neighboring elements is equal to $\lambda/16$.

The forth set under investigation has the following transmit elements: $(2,2)$, $(2,7)$, $(4,4)$, $(4,5)$, $(5,4)$, $(5,5)$, $(7,2)$, and $(7,7)$. Each of $(2,2)$, $(2,7)$, $(7,2)$, and $(7,7)$ elements has $4$ first-order neighboring elements and $4$ second-order neighboring elements, while each of $(4,4)$, $(4,5)$, $(5,4)$, $(5,5)$ has $2$ first-order neighboring passive elements and $3$ second-order neighboring passive elements. As the fourth set has a higher numberof neighboring elements than the first and second sets, the accuracy of the prediction engine for this set should be higher than the one of the first and second sets. Moreover, due to the low correlation of the transmit elements with their second-order neighboring elements as well as the similar correlations between the transmit elements, the performance of the prediction engines of the third and forth sets are expected to be similar.

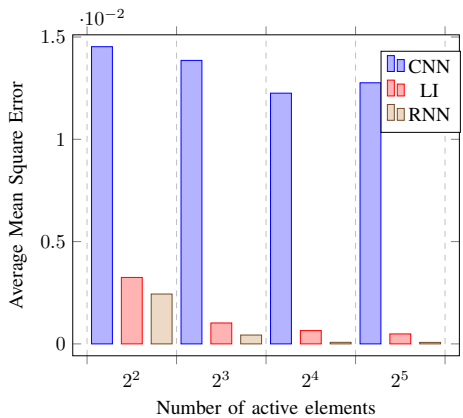
\begin{figure}
	\centering
	\scalebox{0.75}{\input{Angles_compare}}
	\caption{AMSE of the channel phase at different SORIS transmit elements for different deep learning network. In this figure CNN and LI stands for convolution neural network and linear interpolation, respectively.}
	\label{Fig:fig31}
\end{figure}

 {In Fig.~\ref{Fig:fig31}, the AMSE is plotted as a function of the number of active elements for different types of predictors. In more detail, we used a convolution neural network (CNN), a linear interpolator (LI), and an RNN. For the case in which $4$ active elements are used, the active transmit elements are $(1,1)$, $(1,8)$, $(8,1)$ and $(8,8)$. For the case in which $8$ active elements are employed, they are $(1,1)$, $(1,8)$, $(3,3)$, $(3,6)$, $(6,3)$, $(6,6)$, $(8,1)$ and $(8,8)$. For the case in which $16$ active elements are used, the selected active elements are $(2,2)$, $(2,4)$, $(2,6)$, $(2,8)$, $(4,2)$, $(4,4)$, $(4,6)$, $(4,8)$, $(6,2)$, $(6,4)$, $(6,6)$, $(6,8)$, $(8,2)$, $(8,4)$, $(8,6)$, and $(8,8)$. Finally, for the case in which $32$ active elements are employed, the selected active elements are $(1,2)$, $(1,4)$, $(1,6)$, $(1,8)$, $(2,1)$, $(2,3)$, $(2,5)$, $(2,7)$, $(3,2)$, $(3,4)$, $(3,6)$, $(3,8)$, $(4,1)$, $(4,3)$, $(4,5)$, $(4,7)$, $(5,2)$, $(5,4)$, $(5,6)$, $(5,8)$, $(6,1)$, $(6,3)$, $(6,5)$, $(6,7)$, $(7,2)$, $(7,4)$, $(7,6)$, $(7,8)$, $(8,1)$, $(8,3)$, $(8,5)$, and $(8,7)$. As expected, for a given type of predictor, as the number of active transmit elements increases, the AMSE decreases. Additionally, from this figure, it becomes evident that RNN outperforms both CNN and LI. Finally, LI outperforms CNN, in terms of AMSE. These results indicate the importance of investigating and selecting the predictor. A suitable selection is expected to considerably reduce the AMSE.}  

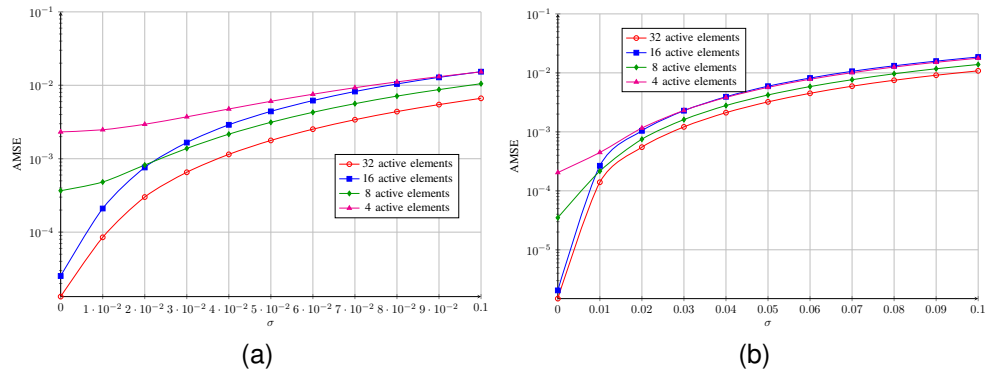
\begin{figure*}
	\centering
    \subfloat[]{\scalebox{0.45}{\input{Angles_predict}}
	\label{Fig:fig32}}
    \subfloat[]{\scalebox{0.45}{\input{abs_predict}}
    \label{Fig:fig33}}
    \caption{AMSE of channel coefficients’ amplitude (a) and phase (b) as a function of the standard deviation of the channel estimator error for different values of channel elements and SNR equals to $-3\,\rm{dB}$.}
    \label{Fig:fig34}
\end{figure*}

 {Fig.~\ref{Fig:fig34} illustrates the impact of the channel estimator error on the performance of the RNN channel predictor. In more detail, the AMSE of channel coefficients’ amplitude and phase are depicted as a function of the standard deviation of the channel estimator error for different values of channel elements. The channel estimator error is modeled as a zero-mean complex Gaussian process with variance equal to $\sigma^2$. We assume that the channel estimator error is due to the receiver's hardware imperfections; thus, its standard deviation is lower than $0.1$, which is the maximum acceptable error vector magnitude according to~\cite{9159653}. From this figure, we observe that for a given number of active elements, as the channel estimator standard deviation increases, the AMSE of the channel predictor also increases. Moreover, for a given $\sigma$, as the number of active elements increases from $4$ to $8$ and then $16$, the AMSE decreases. }

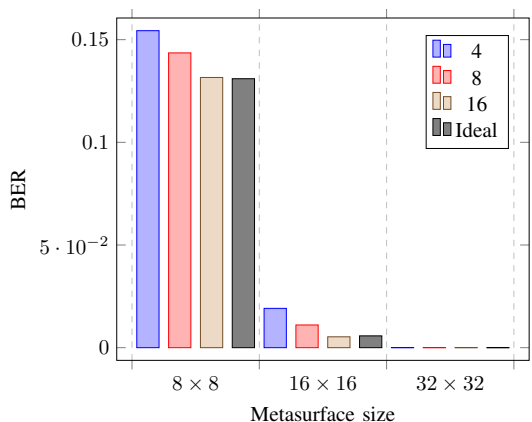
\begin{figure}
	\centering
    {\scalebox{0.8}{{ \input{Bigger_RIS_BER} }}}
    \label{Fig:fig41}
    \caption{BER vs metasurface size for different values of active elements.}
\end{figure}

 {Fig.~12 depicts the BER as a function of the RIS size for different numbers of active elements. As a benchmark, the case with perfect knowledge of CSI at the RIS micro-controller is also plotted. As expected, for a given number of active elements, as the RIS size increases, the diversity gain increases; thus, the BER decreases. Moreover, for a fixed RIS size, as the number of active elements increases, the CSI prediction accuracy increases; as a consequence, the BER decreases. Interestingly, we observe that, regardless of the RIS size for $16$  active elements, SORIS achieves similar results as the ideal case in which the micro-controller has perfect CSI knowledge.}


\subsection{Selection of the SORIS Active Transmit Elements}\label{SS:Guidelines}
From the above observation, it becomes evident that the key parameters that affect both the number of transmit elements as well as their relative position are: \textit{i}) their mutual correlation; and \textit{ii}) the correlations between the transmit elements and the first- and second-order neighboring elements. The key concept lies on selecting transmit elements that provide information for the SORIS areas in which pre-selected transmit elements are incapable to acquire~knowledge. 

Building upon the aforementioned principle, the first transmit element of the SORIS is selected upon the main diagonal of SORIS, i.e., $(k,k)$ with $k$ being a natural number that is lower than $\min(M_h, M_v)$. From our simulations, we observed that the correlation between the $(1,1)$ and $(k,k)$ elements is in the range of $[0.2, 0.4]$. The next transmit element should be $(k,k+K)$ with $k+K\leq M_v$. Parameter $K$ is selected in a manner in which the correlation between $(k,k)$ and $(k, k+K)$ is lower than $0.1$. Intuitively, this means that $(k, k+K)$ provides information about the channel distribution upon SORIS for an area for which $(k,k)$ is unable to acquire any knowledge. The third transmit element is selected in a similar manner and is the $(k+K, k)$, with $k+K\leq N_h$. Each of the next elements that is set as active is one of $(k + n_v\,K, k + n_h\,K)$ with $n_v$ and $n_h$ been naturals and  $k + n_v\,K \leq M_v$ as well as $k + n_h\,K \leq N_h$. In the case of any horizontal or vertical symmetry, i.e., if element with either $(k_1, l_1)$ and $(k_2, l_1)$ with $k_1\neq k_2$ or  $(k_1, l_1)$ and $(k_1, l_2)$ with $l_1\neq l_2$ are selected, then, it is preferable to change the set into $(k_1, l_1)$ and $(k_2, l_1+1)$ or  $(k_1, l_1)$ and $(k_1+1, l_2)$ assuming that $l_1+1\leq M_v$ and  and $k_1+1\leq M_h$.

\section{Conclusions and Future Directions}\label{S:Con}
This paper presented a novel RIS hardware architecture, named SORIS, being able to acquire both the BS-SORIS and SORIS-UE CSI for each of the SORIS elements in almost real-time and with low complexity and moderate channel acquisition latency. To demonstrate the efficiency of the proposed architecture, we presented an ML-empowered channel prediction framework that collects CSI for a small number of active transmit SORIS elements and predicts the channel coefficients at all the other SORIS elements. 
After acquiring the instantaneous CSI for both the BS-SORIS and SORIS-UE channels, the proposed RIS architecture can perform real-time adaptation to improve the data transmission rate.
The efficiency of the SORIS CSI acquisition approach was assessed through Monte Carlo simulations. It was observed that, as the number of transmit elements, and as a consequence the channel acquisition latency increases, the channel prediction accuracy improves. Moreover, it was demonstrated that, as the distance between neighboring SORIS elements decreases, their correlation increases, and thus the accuracy of our channel prediction approach increases.  {The importance of choosing a suitable ML approach in order to minimize the AMSE was also highlighted. The robustness of the channel prediction scheme against partial channel knowledge due to the micro-controller transceiver hardware imperfections was discussed.} Finally, it became evident that a proper selection of the active transmit SORIS elements can minimize the AMSE in the prediction of both the magnitude and phase of the channel coefficients. This motivated the design of a procedure for appropriately selecting the active elements for initial channel estimation.

 {In this work, we assumed that the micro-controller's receiver was ideal. In other words, the impact of hardware imperfections, like in-phase and quadrature imbalance, phase noise, and low noise amplifier non-nonlinearities were neglected. As discussed in several prior papers, including~\cite{9159653,7378527,9026774}, hardware imperfections are expected to significantly degrade the performance of the channel estimator. Motivated by this, as a next step, we aim to quantify the impact of hardware imperfections. In the same direction, DSP approaches that more effectively mitigate the impact of hardware imperfections and improve the channel estimation performance need to be investigated. Another future direction lies in the assessment of the power dissipation of the phase switching layer for different phase switching architectures. Although the power dissipation is expected to be lower or equal to STAR-RIS architecture, a novel phase switching layer approach is expected to have a positive contribution on the energy efficiency of SORIS.} 

 {To further improve the performance of SORIS, in terms of channel prediction error, and energy efficiency, we need to study new approaches from the computing perspective. Specifically, we will steer our attention to } the design of hardware-aware ML-approaches for channel prediction that minimize the energy consumption of the processor.  {As discussed in~\cite{10525226}, such approaches need dedicated hardware, i.e., artificial intelligence accelerators, like neuromorphic computing units, which are coupled with specific ML agent families, such as spiking neural networks. The performance gains in terms of energy efficiency and response time, as well as the assessment of the channel prediction error against conventional approaches, like CNN, RNN, and LI, are directions worth pursuing.}   


\bibliographystyle{IEEEtran}
\bibliography{refs}

\end{document}

%% file: systemModel.tex
  
\tikzset {_jszgvq2xx/.code = {\pgfsetadditionalshadetransform{ \pgftransformshift{\pgfpoint{0 bp } { 0 bp }  }  \pgftransformrotate{-270 }  \pgftransformscale{2 }  }}}
\pgfdeclarehorizontalshading{_u9rx0em9x}{150bp}{rgb(0bp)=(0.8,0.78,0.78);
rgb(37.5bp)=(0.8,0.78,0.78);
rgb(62.5bp)=(0.32,0.31,0.31);
rgb(100bp)=(0.32,0.31,0.31)}
\tikzset{_pkoadjeri/.code = {\pgfsetadditionalshadetransform{\pgftransformshift{\pgfpoint{0 bp } { 0 bp }  }  \pgftransformrotate{-270 }  \pgftransformscale{2 } }}}
\pgfdeclarehorizontalshading{_yiylzqnps} {150bp} {color(0bp)=(transparent!0);
color(37.5bp)=(transparent!0);
color(62.5bp)=(transparent!10);
color(100bp)=(transparent!10) } 
\pgfdeclarefading{_259v3ot5q}{\tikz \fill[shading=_yiylzqnps,_pkoadjeri] (0,0) rectangle (50bp,50bp); } 

  
\tikzset {_wiczeaye1/.code = {\pgfsetadditionalshadetransform{ \pgftransformshift{\pgfpoint{0 bp } { 0 bp }  }  \pgftransformrotate{0 }  \pgftransformscale{2 }  }}}
\pgfdeclarehorizontalshading{_uvtalctwj}{150bp}{rgb(0bp)=(0.97,0.62,0.04);
rgb(37.5bp)=(0.97,0.62,0.04);
rgb(50bp)=(1,1,0);
rgb(62.5bp)=(1,0,0);
rgb(100bp)=(1,0,0)}

  
\tikzset {_zc5fijl4k/.code = {\pgfsetadditionalshadetransform{ \pgftransformshift{\pgfpoint{0 bp } { 0 bp }  }  \pgftransformrotate{0 }  \pgftransformscale{2 }  }}}
\pgfdeclarehorizontalshading{_16kjjrch0}{150bp}{rgb(0bp)=(1,0,0);
rgb(37.5bp)=(1,0,0);
rgb(43.75bp)=(1,1,0);
rgb(50bp)=(0.02,0.76,1);
rgb(56.25bp)=(1,1,0);
rgb(62.5bp)=(1,0,0);
rgb(100bp)=(1,0,0)}

  
\tikzset {_usmc87hnb/.code = {\pgfsetadditionalshadetransform{ \pgftransformshift{\pgfpoint{0 bp } { 0 bp }  }  \pgftransformrotate{0 }  \pgftransformscale{2 }  }}}
\pgfdeclarehorizontalshading{_4e74hdcuv}{150bp}{rgb(0bp)=(0.97,0.62,0.04);
rgb(37.5bp)=(0.97,0.62,0.04);
rgb(50bp)=(1,1,0);
rgb(62.5bp)=(1,0,0);
rgb(100bp)=(1,0,0)}

  
\tikzset {_ujmdoqea4/.code = {\pgfsetadditionalshadetransform{ \pgftransformshift{\pgfpoint{0 bp } { 0 bp }  }  \pgftransformrotate{-180 }  \pgftransformscale{2 }  }}}
\pgfdeclarehorizontalshading{_mtux42d14}{150bp}{rgb(0bp)=(0.93,0.93,0.1);
rgb(37.5bp)=(0.93,0.93,0.1);
rgb(62.5bp)=(0.66,0.43,0.02);
rgb(100bp)=(0.66,0.43,0.02)}

  
\tikzset {_bo5mlfaem/.code = {\pgfsetadditionalshadetransform{ \pgftransformshift{\pgfpoint{0 bp } { 0 bp }  }  \pgftransformrotate{0 }  \pgftransformscale{2 }  }}}
\pgfdeclarehorizontalshading{_imp0t112f}{150bp}{rgb(0bp)=(0.97,0.62,0.04);
rgb(37.5bp)=(0.97,0.62,0.04);
rgb(50bp)=(1,1,0);
rgb(62.5bp)=(1,0,0);
rgb(100bp)=(1,0,0)}

  
\tikzset {_srov33zi1/.code = {\pgfsetadditionalshadetransform{ \pgftransformshift{\pgfpoint{0 bp } { 0 bp }  }  \pgftransformrotate{0 }  \pgftransformscale{2 }  }}}
\pgfdeclarehorizontalshading{_e72myzipk}{150bp}{rgb(0bp)=(0.84,0.83,0.83);
rgb(37.5bp)=(0.84,0.83,0.83);
rgb(62.5bp)=(0,0,0);
rgb(100bp)=(0,0,0)}
\tikzset{_ynz9x5o0i/.code = {\pgfsetadditionalshadetransform{\pgftransformshift{\pgfpoint{0 bp } { 0 bp }  }  \pgftransformrotate{0 }  \pgftransformscale{2 } }}}
\pgfdeclarehorizontalshading{_z5j4128xy} {150bp} {color(0bp)=(transparent!0);
color(37.5bp)=(transparent!0);
color(62.5bp)=(transparent!10);
color(100bp)=(transparent!10) } 
\pgfdeclarefading{_9zuby0o8p}{\tikz \fill[shading=_z5j4128xy,_ynz9x5o0i] (0,0) rectangle (50bp,50bp); } 

  
\tikzset {_w75ocjv63/.code = {\pgfsetadditionalshadetransform{ \pgftransformshift{\pgfpoint{0 bp } { 0 bp }  }  \pgftransformrotate{-90 }  \pgftransformscale{2 }  }}}
\pgfdeclarehorizontalshading{_pnpra1nbf}{150bp}{rgb(0bp)=(1,0,0);
rgb(37.5bp)=(1,0,0);
rgb(43.75bp)=(1,1,0);
rgb(50bp)=(0.02,0.76,1);
rgb(56.25bp)=(1,1,0);
rgb(62.5bp)=(1,0,0);
rgb(100bp)=(1,0,0)}

  
\tikzset {_xz6zbha2e/.code = {\pgfsetadditionalshadetransform{ \pgftransformshift{\pgfpoint{0 bp } { 0 bp }  }  \pgftransformrotate{-270 }  \pgftransformscale{2 }  }}}
\pgfdeclarehorizontalshading{_4bq82r9c8}{150bp}{rgb(0bp)=(0.61,0.61,0.61);
rgb(37.5bp)=(0.61,0.61,0.61);
rgb(62.5bp)=(0,0,0);
rgb(100bp)=(0,0,0)}
\tikzset{_tgyle7hq3/.code = {\pgfsetadditionalshadetransform{\pgftransformshift{\pgfpoint{0 bp } { 0 bp }  }  \pgftransformrotate{-270 }  \pgftransformscale{2 } }}}
\pgfdeclarehorizontalshading{_ys0b6zpo9} {150bp} {color(0bp)=(transparent!0);
color(37.5bp)=(transparent!0);
color(62.5bp)=(transparent!10);
color(100bp)=(transparent!10) } 
\pgfdeclarefading{_b7ie1pk3k}{\tikz \fill[shading=_ys0b6zpo9,_tgyle7hq3] (0,0) rectangle (50bp,50bp); } 

  
\tikzset {_ltfwf60ws/.code = {\pgfsetadditionalshadetransform{ \pgftransformshift{\pgfpoint{0 bp } { 0 bp }  }  \pgftransformrotate{0 }  \pgftransformscale{2 }  }}}
\pgfdeclarehorizontalshading{_kymnuo23h}{150bp}{rgb(0bp)=(0.97,0.62,0.04);
rgb(37.5bp)=(0.97,0.62,0.04);
rgb(50bp)=(1,1,0);
rgb(62.5bp)=(1,0,0);
rgb(100bp)=(1,0,0)}

  
\tikzset {_vlh58s86n/.code = {\pgfsetadditionalshadetransform{ \pgftransformshift{\pgfpoint{0 bp } { 0 bp }  }  \pgftransformrotate{0 }  \pgftransformscale{2 }  }}}
\pgfdeclarehorizontalshading{_m82snt6qh}{150bp}{rgb(0bp)=(0.97,0.62,0.04);
rgb(37.5bp)=(0.97,0.62,0.04);
rgb(50bp)=(1,1,0);
rgb(62.5bp)=(1,0,0);
rgb(100bp)=(1,0,0)}

  
\tikzset {_7d6bp3eak/.code = {\pgfsetadditionalshadetransform{ \pgftransformshift{\pgfpoint{0 bp } { 0 bp }  }  \pgftransformrotate{0 }  \pgftransformscale{2 }  }}}
\pgfdeclarehorizontalshading{_0f8c73vru}{150bp}{rgb(0bp)=(0.49,0.83,0.13);
rgb(37.5bp)=(0.49,0.83,0.13);
rgb(50bp)=(0.49,0.83,0.13);
rgb(62.5bp)=(0.1,0.18,0);
rgb(100bp)=(0.1,0.18,0)}

  
\tikzset {_skk7kfusv/.code = {\pgfsetadditionalshadetransform{ \pgftransformshift{\pgfpoint{0 bp } { 0 bp }  }  \pgftransformrotate{0 }  \pgftransformscale{2 }  }}}
\pgfdeclarehorizontalshading{_in1rcz224}{150bp}{rgb(0bp)=(0.49,0.83,0.13);
rgb(37.5bp)=(0.49,0.83,0.13);
rgb(50bp)=(0.49,0.83,0.13);
rgb(62.5bp)=(0.1,0.18,0);
rgb(100bp)=(0.1,0.18,0)}

  
\tikzset {_660gnrfta/.code = {\pgfsetadditionalshadetransform{ \pgftransformshift{\pgfpoint{0 bp } { 0 bp }  }  \pgftransformrotate{0 }  \pgftransformscale{2 }  }}}
\pgfdeclarehorizontalshading{_0rccx2sqs}{150bp}{rgb(0bp)=(0.49,0.83,0.13);
rgb(37.5bp)=(0.49,0.83,0.13);
rgb(50bp)=(0.49,0.83,0.13);
rgb(62.5bp)=(0.1,0.18,0);
rgb(100bp)=(0.1,0.18,0)}
\tikzset{every picture/.style={line width=0.75pt}} 

\begin{tikzpicture}[x=0.75pt,y=0.75pt,yscale=-1,xscale=1]

\path  [shading=_u9rx0em9x,_jszgvq2xx,path fading= _259v3ot5q ,fading transform={xshift=2}] (286,44.24) -- (299.53,30.71) -- (387.41,30.71) -- (387.41,46.32) -- (373.88,59.84) -- (286,59.84) -- cycle ; 
 \draw   (286,44.24) -- (299.53,30.71) -- (387.41,30.71) -- (387.41,46.32) -- (373.88,59.84) -- (286,59.84) -- cycle ; 
 \draw   (387.41,30.71) -- (373.88,44.24) -- (286,44.24) ; \draw   (373.88,44.24) -- (373.88,59.84) ;

\draw  [draw opacity=0][shading=_uvtalctwj,_wiczeaye1] (317.5,35) .. controls (334.5,64) and (412.05,174) .. (424.5,186) .. controls (436.95,198) and (319.13,25.3) .. (307.13,14.3) .. controls (295.13,3.3) and (300.5,6) .. (317.5,35) -- cycle ;
\draw  [draw opacity=0][shading=_16kjjrch0,_zc5fijl4k] (411.84,88.71) .. controls (412.38,88.71) and (412.81,84.87) .. (412.81,80.13) .. controls (412.81,75.39) and (412.38,71.54) .. (411.84,71.54) .. controls (411.3,71.54) and (410.87,67.7) .. (410.87,62.96) .. controls (410.87,58.22) and (411.3,54.37) .. (411.84,54.37) -- (419.61,54.37) .. controls (419.07,54.37) and (418.64,58.22) .. (418.64,62.96) .. controls (418.64,67.7) and (419.07,71.54) .. (419.61,71.54) .. controls (420.14,71.54) and (420.58,75.39) .. (420.58,80.13) .. controls (420.58,84.87) and (420.14,88.71) .. (419.61,88.71) -- cycle ;
\draw  [draw opacity=0][shading=_4e74hdcuv,_usmc87hnb] (265.51,53.33) .. controls (243.4,90.4) and (241.5,116.5) .. (228,132) .. controls (214.5,147.5) and (284.65,38.9) .. (297.51,19.58) .. controls (310.38,0.25) and (287.63,16.25) .. (265.51,53.33) -- cycle ;
\draw  [fill={rgb, 255:red, 141; green, 141; blue, 141 }  ,fill opacity=1 ] (198,81) -- (465.75,81) -- (465.75,231) -- (198,231) -- cycle ;
\path  [shading=_mtux42d14,_ujmdoqea4] (206,88.36) -- (458,88.36) -- (458,225) -- (206,225) -- cycle ; 
 \draw   (206,88.36) -- (458,88.36) -- (458,225) -- (206,225) -- cycle ; 

\draw  [fill={rgb, 255:red, 128; green, 128; blue, 128 }  ,fill opacity=1 ] (209,93) -- (456,93) -- (456,221) -- (209,221) -- cycle ;
\draw  [fill={rgb, 255:red, 95; green, 60; blue, 2 }  ,fill opacity=1 ] (213,96) -- (228,96) -- (228,112) -- (213,112) -- cycle ;
\draw  [fill={rgb, 255:red, 95; green, 60; blue, 2 }  ,fill opacity=1 ] (232,96) -- (247,96) -- (247,112) -- (232,112) -- cycle ;
\draw  [fill={rgb, 255:red, 95; green, 60; blue, 2 }  ,fill opacity=1 ] (250,96) -- (265,96) -- (265,112) -- (250,112) -- cycle ;
\draw  [fill={rgb, 255:red, 95; green, 60; blue, 2 }  ,fill opacity=1 ] (268,96) -- (283,96) -- (283,112) -- (268,112) -- cycle ;
\draw  [fill={rgb, 255:red, 95; green, 60; blue, 2 }  ,fill opacity=1 ] (287,96) -- (302,96) -- (302,112) -- (287,112) -- cycle ;
\draw  [fill={rgb, 255:red, 95; green, 60; blue, 2 }  ,fill opacity=1 ] (306,96) -- (321,96) -- (321,112) -- (306,112) -- cycle ;
\draw  [fill={rgb, 255:red, 95; green, 60; blue, 2 }  ,fill opacity=1 ] (324,96) -- (339,96) -- (339,112) -- (324,112) -- cycle ;
\draw  [fill={rgb, 255:red, 95; green, 60; blue, 2 }  ,fill opacity=1 ] (342,96) -- (357,96) -- (357,112) -- (342,112) -- cycle ;
\draw  [fill={rgb, 255:red, 95; green, 60; blue, 2 }  ,fill opacity=1 ] (362,96) -- (377,96) -- (377,112) -- (362,112) -- cycle ;
\draw  [fill={rgb, 255:red, 95; green, 60; blue, 2 }  ,fill opacity=1 ] (381,96) -- (396,96) -- (396,112) -- (381,112) -- cycle ;
\draw  [fill={rgb, 255:red, 95; green, 60; blue, 2 }  ,fill opacity=1 ] (399,96) -- (414,96) -- (414,112) -- (399,112) -- cycle ;
\draw  [fill={rgb, 255:red, 95; green, 60; blue, 2 }  ,fill opacity=1 ] (417,96) -- (432,96) -- (432,112) -- (417,112) -- cycle ;
\draw  [fill={rgb, 255:red, 95; green, 60; blue, 2 }  ,fill opacity=1 ] (436,96) -- (451,96) -- (451,112) -- (436,112) -- cycle ;

\draw  [fill={rgb, 255:red, 95; green, 60; blue, 2 }  ,fill opacity=1 ] (213,116) -- (228,116) -- (228,132) -- (213,132) -- cycle ;
\draw  [fill={rgb, 255:red, 208; green, 2; blue, 27 }  ,fill opacity=1 ] (232,116) -- (247,116) -- (247,132) -- (232,132) -- cycle ;
\draw  [fill={rgb, 255:red, 95; green, 60; blue, 2 }  ,fill opacity=1 ] (250,116) -- (265,116) -- (265,132) -- (250,132) -- cycle ;
\draw  [fill={rgb, 255:red, 95; green, 60; blue, 2 }  ,fill opacity=1 ] (268,116) -- (283,116) -- (283,132) -- (268,132) -- cycle ;
\draw  [fill={rgb, 255:red, 95; green, 60; blue, 2 }  ,fill opacity=1 ] (287,116) -- (302,116) -- (302,132) -- (287,132) -- cycle ;
\draw  [fill={rgb, 255:red, 95; green, 60; blue, 2 }  ,fill opacity=1 ] (306,116) -- (321,116) -- (321,132) -- (306,132) -- cycle ;
\draw  [fill={rgb, 255:red, 95; green, 60; blue, 2 }  ,fill opacity=1 ] (324,116) -- (339,116) -- (339,132) -- (324,132) -- cycle ;
\draw  [fill={rgb, 255:red, 95; green, 60; blue, 2 }  ,fill opacity=1 ] (342,116) -- (357,116) -- (357,132) -- (342,132) -- cycle ;
\draw  [fill={rgb, 255:red, 95; green, 60; blue, 2 }  ,fill opacity=1 ] (362,116) -- (377,116) -- (377,132) -- (362,132) -- cycle ;
\draw  [fill={rgb, 255:red, 95; green, 60; blue, 2 }  ,fill opacity=1 ] (381,116) -- (396,116) -- (396,132) -- (381,132) -- cycle ;
\draw  [fill={rgb, 255:red, 95; green, 60; blue, 2 }  ,fill opacity=1 ] (399,116) -- (414,116) -- (414,132) -- (399,132) -- cycle ;
\draw  [fill={rgb, 255:red, 95; green, 60; blue, 2 }  ,fill opacity=1 ] (417,116) -- (432,116) -- (432,132) -- (417,132) -- cycle ;
\draw  [fill={rgb, 255:red, 95; green, 60; blue, 2 }  ,fill opacity=1 ] (436,116) -- (451,116) -- (451,132) -- (436,132) -- cycle ;
\draw  [fill={rgb, 255:red, 95; green, 60; blue, 2 }  ,fill opacity=1 ] (213,138) -- (228,138) -- (228,154) -- (213,154) -- cycle ;
\draw  [fill={rgb, 255:red, 95; green, 60; blue, 2 }  ,fill opacity=1 ] (232,138) -- (247,138) -- (247,154) -- (232,154) -- cycle ;
\draw  [fill={rgb, 255:red, 95; green, 60; blue, 2 }  ,fill opacity=1 ] (250,138) -- (265,138) -- (265,154) -- (250,154) -- cycle ;
\draw  [fill={rgb, 255:red, 95; green, 60; blue, 2 }  ,fill opacity=1 ] (268,138) -- (283,138) -- (283,154) -- (268,154) -- cycle ;
\draw  [fill={rgb, 255:red, 95; green, 60; blue, 2 }  ,fill opacity=1 ] (287,138) -- (302,138) -- (302,154) -- (287,154) -- cycle ;
\draw  [fill={rgb, 255:red, 95; green, 60; blue, 2 }  ,fill opacity=1 ] (306,138) -- (321,138) -- (321,154) -- (306,154) -- cycle ;
\draw  [fill={rgb, 255:red, 95; green, 60; blue, 2 }  ,fill opacity=1 ] (324,138) -- (339,138) -- (339,154) -- (324,154) -- cycle ;
\draw  [fill={rgb, 255:red, 95; green, 60; blue, 2 }  ,fill opacity=1 ] (342,138) -- (357,138) -- (357,154) -- (342,154) -- cycle ;
\draw  [fill={rgb, 255:red, 95; green, 60; blue, 2 }  ,fill opacity=1 ] (362,138) -- (377,138) -- (377,154) -- (362,154) -- cycle ;
\draw  [fill={rgb, 255:red, 95; green, 60; blue, 2 }  ,fill opacity=1 ] (381,138) -- (396,138) -- (396,154) -- (381,154) -- cycle ;
\draw  [fill={rgb, 255:red, 95; green, 60; blue, 2 }  ,fill opacity=1 ] (399,138) -- (414,138) -- (414,154) -- (399,154) -- cycle ;
\draw  [fill={rgb, 255:red, 95; green, 60; blue, 2 }  ,fill opacity=1 ] (417,138) -- (432,138) -- (432,154) -- (417,154) -- cycle ;
\draw  [fill={rgb, 255:red, 95; green, 60; blue, 2 }  ,fill opacity=1 ] (436,138) -- (451,138) -- (451,154) -- (436,154) -- cycle ;

\draw  [fill={rgb, 255:red, 95; green, 60; blue, 2 }  ,fill opacity=1 ] (213,158) -- (228,158) -- (228,174) -- (213,174) -- cycle ;
\draw  [fill={rgb, 255:red, 95; green, 60; blue, 2 }  ,fill opacity=1 ] (232,158) -- (247,158) -- (247,174) -- (232,174) -- cycle ;
\draw  [fill={rgb, 255:red, 95; green, 60; blue, 2 }  ,fill opacity=1 ] (250,158) -- (265,158) -- (265,174) -- (250,174) -- cycle ;
\draw  [fill={rgb, 255:red, 95; green, 60; blue, 2 }  ,fill opacity=1 ] (268,158) -- (283,158) -- (283,174) -- (268,174) -- cycle ;
\draw  [fill={rgb, 255:red, 95; green, 60; blue, 2 }  ,fill opacity=1 ] (287,158) -- (302,158) -- (302,174) -- (287,174) -- cycle ;
\draw  [fill={rgb, 255:red, 95; green, 60; blue, 2 }  ,fill opacity=1 ] (306,158) -- (321,158) -- (321,174) -- (306,174) -- cycle ;
\draw  [fill={rgb, 255:red, 95; green, 60; blue, 2 }  ,fill opacity=1 ] (324,158) -- (339,158) -- (339,174) -- (324,174) -- cycle ;
\draw  [fill={rgb, 255:red, 95; green, 60; blue, 2 }  ,fill opacity=1 ] (342,158) -- (357,158) -- (357,174) -- (342,174) -- cycle ;
\draw  [fill={rgb, 255:red, 95; green, 60; blue, 2 }  ,fill opacity=1 ] (362,158) -- (377,158) -- (377,174) -- (362,174) -- cycle ;
\draw  [fill={rgb, 255:red, 95; green, 60; blue, 2 }  ,fill opacity=1 ] (381,158) -- (396,158) -- (396,174) -- (381,174) -- cycle ;
\draw  [fill={rgb, 255:red, 95; green, 60; blue, 2 }  ,fill opacity=1 ] (399,158) -- (414,158) -- (414,174) -- (399,174) -- cycle ;
\draw  [fill={rgb, 255:red, 95; green, 60; blue, 2 }  ,fill opacity=1 ] (417,158) -- (432,158) -- (432,174) -- (417,174) -- cycle ;
\draw  [fill={rgb, 255:red, 95; green, 60; blue, 2 }  ,fill opacity=1 ] (436,158) -- (451,158) -- (451,174) -- (436,174) -- cycle ;

\draw  [fill={rgb, 255:red, 95; green, 60; blue, 2 }  ,fill opacity=1 ] (213,178) -- (228,178) -- (228,194) -- (213,194) -- cycle ;
\draw  [fill={rgb, 255:red, 95; green, 60; blue, 2 }  ,fill opacity=1 ] (232,178) -- (247,178) -- (247,194) -- (232,194) -- cycle ;
\draw  [fill={rgb, 255:red, 95; green, 60; blue, 2 }  ,fill opacity=1 ] (250,178) -- (265,178) -- (265,194) -- (250,194) -- cycle ;
\draw  [fill={rgb, 255:red, 95; green, 60; blue, 2 }  ,fill opacity=1 ] (268,178) -- (283,178) -- (283,194) -- (268,194) -- cycle ;
\draw  [fill={rgb, 255:red, 95; green, 60; blue, 2 }  ,fill opacity=1 ] (287,178) -- (302,178) -- (302,194) -- (287,194) -- cycle ;
\draw  [fill={rgb, 255:red, 95; green, 60; blue, 2 }  ,fill opacity=1 ] (306,178) -- (321,178) -- (321,194) -- (306,194) -- cycle ;
\draw  [fill={rgb, 255:red, 95; green, 60; blue, 2 }  ,fill opacity=1 ] (324,178) -- (339,178) -- (339,194) -- (324,194) -- cycle ;
\draw  [fill={rgb, 255:red, 95; green, 60; blue, 2 }  ,fill opacity=1 ] (342,178) -- (357,178) -- (357,194) -- (342,194) -- cycle ;
\draw  [fill={rgb, 255:red, 95; green, 60; blue, 2 }  ,fill opacity=1 ] (362,178) -- (377,178) -- (377,194) -- (362,194) -- cycle ;
\draw  [fill={rgb, 255:red, 95; green, 60; blue, 2 }  ,fill opacity=1 ] (381,178) -- (396,178) -- (396,194) -- (381,194) -- cycle ;
\draw  [fill={rgb, 255:red, 95; green, 60; blue, 2 }  ,fill opacity=1 ] (399,178) -- (414,178) -- (414,194) -- (399,194) -- cycle ;
\draw  [fill={rgb, 255:red, 208; green, 2; blue, 27 }  ,fill opacity=1 ] (417,178) -- (432,178) -- (432,194) -- (417,194) -- cycle ;
\draw  [fill={rgb, 255:red, 95; green, 60; blue, 2 }  ,fill opacity=1 ] (436,178) -- (451,178) -- (451,194) -- (436,194) -- cycle ;
\draw  [fill={rgb, 255:red, 95; green, 60; blue, 2 }  ,fill opacity=1 ] (213,198) -- (228,198) -- (228,214) -- (213,214) -- cycle ;
\draw  [fill={rgb, 255:red, 95; green, 60; blue, 2 }  ,fill opacity=1 ] (232,198) -- (247,198) -- (247,214) -- (232,214) -- cycle ;
\draw  [fill={rgb, 255:red, 95; green, 60; blue, 2 }  ,fill opacity=1 ] (250,198) -- (265,198) -- (265,214) -- (250,214) -- cycle ;
\draw  [fill={rgb, 255:red, 95; green, 60; blue, 2 }  ,fill opacity=1 ] (268,198) -- (283,198) -- (283,214) -- (268,214) -- cycle ;
\draw  [fill={rgb, 255:red, 95; green, 60; blue, 2 }  ,fill opacity=1 ] (287,198) -- (302,198) -- (302,214) -- (287,214) -- cycle ;
\draw  [fill={rgb, 255:red, 95; green, 60; blue, 2 }  ,fill opacity=1 ] (306,198) -- (321,198) -- (321,214) -- (306,214) -- cycle ;
\draw  [fill={rgb, 255:red, 95; green, 60; blue, 2 }  ,fill opacity=1 ] (324,198) -- (339,198) -- (339,214) -- (324,214) -- cycle ;
\draw  [fill={rgb, 255:red, 95; green, 60; blue, 2 }  ,fill opacity=1 ] (342,198) -- (357,198) -- (357,214) -- (342,214) -- cycle ;
\draw  [fill={rgb, 255:red, 95; green, 60; blue, 2 }  ,fill opacity=1 ] (362,198) -- (377,198) -- (377,214) -- (362,214) -- cycle ;
\draw  [fill={rgb, 255:red, 95; green, 60; blue, 2 }  ,fill opacity=1 ] (381,198) -- (396,198) -- (396,214) -- (381,214) -- cycle ;
\draw  [fill={rgb, 255:red, 95; green, 60; blue, 2 }  ,fill opacity=1 ] (399,198) -- (414,198) -- (414,214) -- (399,214) -- cycle ;
\draw  [fill={rgb, 255:red, 95; green, 60; blue, 2 }  ,fill opacity=1 ] (417,198) -- (432,198) -- (432,214) -- (417,214) -- cycle ;
\draw  [fill={rgb, 255:red, 95; green, 60; blue, 2 }  ,fill opacity=1 ] (436,198) -- (451,198) -- (451,214) -- (436,214) -- cycle ;

\draw  [dash pattern={on 4.5pt off 4.5pt}]  (132,119) -- (229,123.9) ;
\draw [shift={(231,124)}, rotate = 182.89] [color={rgb, 255:red, 0; green, 0; blue, 0 }  ][line width=0.75]    (10.93,-3.29) .. controls (6.95,-1.4) and (3.31,-0.3) .. (0,0) .. controls (3.31,0.3) and (6.95,1.4) .. (10.93,3.29)   ;
\draw  [fill={rgb, 255:red, 240; green, 60; blue, 81 }  ,fill opacity=1 ] (40.5,427.45) -- (40.5,456.55) .. controls (40.5,458.45) and (38.37,460) .. (35.75,460) .. controls (33.13,460) and (31,458.45) .. (31,456.55) -- (31,427.45) .. controls (31,425.55) and (33.13,424) .. (35.75,424) .. controls (38.37,424) and (40.5,425.55) .. (40.5,427.45) .. controls (40.5,429.36) and (38.37,430.91) .. (35.75,430.91) .. controls (33.13,430.91) and (31,429.36) .. (31,427.45) ;
\draw  [fill={rgb, 255:red, 221; green, 233; blue, 247 }  ,fill opacity=1 ] (40.5,398.36) -- (40.5,427.45) .. controls (40.5,429.36) and (38.37,430.91) .. (35.75,430.91) .. controls (33.13,430.91) and (31,429.36) .. (31,427.45) -- (31,398.36) .. controls (31,396.46) and (33.13,394.91) .. (35.75,394.91) .. controls (38.37,394.91) and (40.5,396.46) .. (40.5,398.36) .. controls (40.5,400.27) and (38.37,401.82) .. (35.75,401.82) .. controls (33.13,401.82) and (31,400.27) .. (31,398.36) ;
\draw  [fill={rgb, 255:red, 240; green, 60; blue, 81 }  ,fill opacity=1 ] (40.5,369.27) -- (40.5,398.36) .. controls (40.5,400.27) and (38.37,401.82) .. (35.75,401.82) .. controls (33.13,401.82) and (31,400.27) .. (31,398.36) -- (31,369.27) .. controls (31,367.36) and (33.13,365.82) .. (35.75,365.82) .. controls (38.37,365.82) and (40.5,367.36) .. (40.5,369.27) .. controls (40.5,371.18) and (38.37,372.73) .. (35.75,372.73) .. controls (33.13,372.73) and (31,371.18) .. (31,369.27) ;
\draw  [fill={rgb, 255:red, 221; green, 233; blue, 247 }  ,fill opacity=1 ] (40.5,340.18) -- (40.5,369.27) .. controls (40.5,371.18) and (38.37,372.73) .. (35.75,372.73) .. controls (33.13,372.73) and (31,371.18) .. (31,369.27) -- (31,340.18) .. controls (31,338.27) and (33.13,336.73) .. (35.75,336.73) .. controls (38.37,336.73) and (40.5,338.27) .. (40.5,340.18) .. controls (40.5,342.09) and (38.37,343.64) .. (35.75,343.64) .. controls (33.13,343.64) and (31,342.09) .. (31,340.18) ;
\draw  [fill={rgb, 255:red, 240; green, 60; blue, 81 }  ,fill opacity=1 ] (40.5,311.09) -- (40.5,340.18) .. controls (40.5,342.09) and (38.37,343.64) .. (35.75,343.64) .. controls (33.13,343.64) and (31,342.09) .. (31,340.18) -- (31,311.09) .. controls (31,309.18) and (33.13,307.64) .. (35.75,307.64) .. controls (38.37,307.64) and (40.5,309.18) .. (40.5,311.09) .. controls (40.5,313) and (38.37,314.55) .. (35.75,314.55) .. controls (33.13,314.55) and (31,313) .. (31,311.09) ;

\draw [color={rgb, 255:red, 9; green, 61; blue, 122 }  ,draw opacity=1 ][line width=1.5]    (16.5,417) -- (55,417) ;
\draw [color={rgb, 255:red, 9; green, 61; blue, 122 }  ,draw opacity=1 ][line width=1.5]    (17.5,417) -- (17.5,410) ;
\draw [color={rgb, 255:red, 9; green, 61; blue, 122 }  ,draw opacity=1 ][line width=1.5]    (54,417) -- (54,410) ;
\draw [color={rgb, 255:red, 9; green, 61; blue, 122 }  ,draw opacity=1 ][line width=0.75]    (22.13,417) -- (17.5,410) ;
\draw [color={rgb, 255:red, 9; green, 61; blue, 122 }  ,draw opacity=1 ][line width=0.75]    (27.13,409.75) -- (22.13,417) ;
\draw [color={rgb, 255:red, 9; green, 61; blue, 122 }  ,draw opacity=1 ][line width=0.75]    (30.75,416.75) -- (26.13,409.75) ;
\draw [color={rgb, 255:red, 9; green, 61; blue, 122 }  ,draw opacity=1 ][line width=0.75]    (35.75,409.5) -- (30.75,416.75) ;
\draw [color={rgb, 255:red, 9; green, 61; blue, 122 }  ,draw opacity=1 ][line width=0.75]    (40.13,416.75) -- (35.5,409.75) ;
\draw [color={rgb, 255:red, 9; green, 61; blue, 122 }  ,draw opacity=1 ][line width=0.75]    (45.13,409.5) -- (40.13,416.75) ;
\draw [color={rgb, 255:red, 9; green, 61; blue, 122 }  ,draw opacity=1 ][line width=0.75]    (48.75,416.5) -- (44.13,409.5) ;
\draw [color={rgb, 255:red, 9; green, 61; blue, 122 }  ,draw opacity=1 ][line width=0.75]    (53.75,409.25) -- (48.75,416.5) ;

\draw [color={rgb, 255:red, 9; green, 61; blue, 122 }  ,draw opacity=1 ][line width=1.5]    (16.5,410) -- (55,410) ;

\draw [color={rgb, 255:red, 9; green, 61; blue, 122 }  ,draw opacity=1 ][line width=1.5]    (16.25,394) -- (54.75,394) ;
\draw [color={rgb, 255:red, 9; green, 61; blue, 122 }  ,draw opacity=1 ][line width=1.5]    (17.25,394) -- (17.25,387) ;
\draw [color={rgb, 255:red, 9; green, 61; blue, 122 }  ,draw opacity=1 ][line width=1.5]    (53.75,394) -- (53.75,387) ;
\draw [color={rgb, 255:red, 9; green, 61; blue, 122 }  ,draw opacity=1 ][line width=0.75]    (21.88,394) -- (17.25,387) ;
\draw [color={rgb, 255:red, 9; green, 61; blue, 122 }  ,draw opacity=1 ][line width=0.75]    (26.88,386.75) -- (21.88,394) ;
\draw [color={rgb, 255:red, 9; green, 61; blue, 122 }  ,draw opacity=1 ][line width=0.75]    (30.5,393.75) -- (25.88,386.75) ;
\draw [color={rgb, 255:red, 9; green, 61; blue, 122 }  ,draw opacity=1 ][line width=0.75]    (35.5,386.5) -- (30.5,393.75) ;
\draw [color={rgb, 255:red, 9; green, 61; blue, 122 }  ,draw opacity=1 ][line width=0.75]    (39.88,393.75) -- (35.25,386.75) ;
\draw [color={rgb, 255:red, 9; green, 61; blue, 122 }  ,draw opacity=1 ][line width=0.75]    (44.88,386.5) -- (39.88,393.75) ;
\draw [color={rgb, 255:red, 9; green, 61; blue, 122 }  ,draw opacity=1 ][line width=0.75]    (48.5,393.5) -- (43.88,386.5) ;
\draw [color={rgb, 255:red, 9; green, 61; blue, 122 }  ,draw opacity=1 ][line width=0.75]    (53.5,386.25) -- (48.5,393.5) ;

\draw [color={rgb, 255:red, 9; green, 61; blue, 122 }  ,draw opacity=1 ][line width=1.5]    (16.25,387) -- (54.75,387) ;

\draw  [color={rgb, 255:red, 3; green, 50; blue, 105 }  ,draw opacity=1 ][fill={rgb, 255:red, 126; green, 180; blue, 243 }  ,fill opacity=1 ][line width=0.75]  (28.4,373.36) .. controls (28.8,370.9) and (30.11,369.08) .. (31.34,369.28) .. controls (32.56,369.48) and (33.23,371.63) .. (32.83,374.08) .. controls (32.44,376.53) and (31.12,378.35) .. (29.89,378.16) .. controls (28.67,377.96) and (28,375.81) .. (28.4,373.36) -- cycle ;
\draw  [color={rgb, 255:red, 3; green, 50; blue, 105 }  ,draw opacity=1 ][fill={rgb, 255:red, 126; green, 180; blue, 243 }  ,fill opacity=1 ][line width=0.75]  (39.9,373.86) .. controls (39.5,371.28) and (40.17,369.34) .. (41.4,369.54) .. controls (42.62,369.74) and (43.94,372) .. (44.33,374.58) .. controls (44.73,377.16) and (44.06,379.09) .. (42.84,378.89) .. controls (41.61,378.69) and (40.29,376.44) .. (39.9,373.86) -- cycle ;
\draw    (32.63,372.5) -- (39.13,372.25) ;

\draw [color={rgb, 255:red, 9; green, 61; blue, 122 }  ,draw opacity=1 ][line width=1.5]    (16.5,343.5) -- (55,343.5) ;
\draw [color={rgb, 255:red, 9; green, 61; blue, 122 }  ,draw opacity=1 ][line width=1.5]    (17.5,343.5) -- (17.5,336.5) ;
\draw [color={rgb, 255:red, 9; green, 61; blue, 122 }  ,draw opacity=1 ][line width=1.5]    (54,343.5) -- (54,336.5) ;
\draw [color={rgb, 255:red, 9; green, 61; blue, 122 }  ,draw opacity=1 ][line width=0.75]    (22.13,343.5) -- (17.5,336.5) ;
\draw [color={rgb, 255:red, 9; green, 61; blue, 122 }  ,draw opacity=1 ][line width=0.75]    (27.13,336.25) -- (22.13,343.5) ;
\draw [color={rgb, 255:red, 9; green, 61; blue, 122 }  ,draw opacity=1 ][line width=0.75]    (30.75,343.25) -- (26.13,336.25) ;
\draw [color={rgb, 255:red, 9; green, 61; blue, 122 }  ,draw opacity=1 ][line width=0.75]    (35.75,336) -- (30.75,343.25) ;
\draw [color={rgb, 255:red, 9; green, 61; blue, 122 }  ,draw opacity=1 ][line width=0.75]    (40.13,343.25) -- (35.5,336.25) ;
\draw [color={rgb, 255:red, 9; green, 61; blue, 122 }  ,draw opacity=1 ][line width=0.75]    (45.13,336) -- (40.13,343.25) ;
\draw [color={rgb, 255:red, 9; green, 61; blue, 122 }  ,draw opacity=1 ][line width=0.75]    (48.75,343) -- (44.13,336) ;
\draw [color={rgb, 255:red, 9; green, 61; blue, 122 }  ,draw opacity=1 ][line width=0.75]    (53.75,335.75) -- (48.75,343) ;

\draw [color={rgb, 255:red, 9; green, 61; blue, 122 }  ,draw opacity=1 ][line width=1.5]    (16.5,336.5) -- (55,336.5) ;

\draw  [color={rgb, 255:red, 95; green, 129; blue, 170 }  ,draw opacity=1 ][fill={rgb, 255:red, 158; green, 195; blue, 238 }  ,fill opacity=1 ] (52.25,327.38) -- (55.25,327.38) -- (55.25,344.13) -- (52.25,344.13) -- cycle ;
\draw  [color={rgb, 255:red, 95; green, 129; blue, 170 }  ,draw opacity=1 ][fill={rgb, 255:red, 158; green, 195; blue, 238 }  ,fill opacity=1 ] (34.25,327.63) -- (37.25,327.63) -- (37.25,344.38) -- (34.25,344.38) -- cycle ;
\draw  [color={rgb, 255:red, 95; green, 129; blue, 170 }  ,draw opacity=1 ][fill={rgb, 255:red, 158; green, 195; blue, 238 }  ,fill opacity=1 ] (16.5,326.75) -- (19.5,326.75) -- (19.5,343.5) -- (16.5,343.5) -- cycle ;
\draw  [dash pattern={on 4.5pt off 4.5pt}]  (67.38,368) -- (46.26,374.03) ;
\draw [shift={(44.33,374.58)}, rotate = 344.06] [color={rgb, 255:red, 0; green, 0; blue, 0 }  ][line width=0.75]    (10.93,-3.29) .. controls (6.95,-1.4) and (3.31,-0.3) .. (0,0) .. controls (3.31,0.3) and (6.95,1.4) .. (10.93,3.29)   ;
\draw  [dash pattern={on 4.5pt off 4.5pt}]  (74.38,349.5) -- (57.18,344.67) ;
\draw [shift={(55.25,344.13)}, rotate = 15.7] [color={rgb, 255:red, 0; green, 0; blue, 0 }  ][line width=0.75]    (10.93,-3.29) .. controls (6.95,-1.4) and (3.31,-0.3) .. (0,0) .. controls (3.31,0.3) and (6.95,1.4) .. (10.93,3.29)   ;
\draw  [draw opacity=0][shading=_imp0t112f,_bo5mlfaem] (149.63,219.5) .. controls (123.38,249) and (70.13,318.25) .. (56.63,333.75) .. controls (43.13,349.25) and (236.5,138) .. (239.5,124) .. controls (242.5,110) and (175.88,190) .. (149.63,219.5) -- cycle ;
\path  [shading=_e72myzipk,_srov33zi1,path fading= _9zuby0o8p ,fading transform={xshift=2}] (308.85,13.79) -- (308.85,35.03) .. controls (308.85,35.31) and (308.08,35.54) .. (307.13,35.54) .. controls (306.17,35.54) and (305.4,35.31) .. (305.4,35.03) -- (305.4,13.79) .. controls (305.4,13.5) and (306.17,13.27) .. (307.13,13.27) .. controls (308.08,13.27) and (308.85,13.5) .. (308.85,13.79) .. controls (308.85,14.07) and (308.08,14.3) .. (307.13,14.3) .. controls (306.17,14.3) and (305.4,14.07) .. (305.4,13.79) ; 
 \draw   (308.85,13.79) -- (308.85,35.03) .. controls (308.85,35.31) and (308.08,35.54) .. (307.13,35.54) .. controls (306.17,35.54) and (305.4,35.31) .. (305.4,35.03) -- (305.4,13.79) .. controls (305.4,13.5) and (306.17,13.27) .. (307.13,13.27) .. controls (308.08,13.27) and (308.85,13.5) .. (308.85,13.79) .. controls (308.85,14.07) and (308.08,14.3) .. (307.13,14.3) .. controls (306.17,14.3) and (305.4,14.07) .. (305.4,13.79) ; 

\draw  [draw opacity=0][shading=_pnpra1nbf,_w75ocjv63] (380.85,43.05) .. controls (380.85,43.59) and (383.33,44.02) .. (386.38,44.02) .. controls (389.43,44.02) and (391.91,43.59) .. (391.91,43.05) .. controls (391.91,42.51) and (394.38,42.08) .. (397.44,42.08) .. controls (400.49,42.08) and (402.96,42.51) .. (402.96,43.05) -- (402.96,50.82) .. controls (402.96,50.28) and (400.49,49.85) .. (397.44,49.85) .. controls (394.38,49.85) and (391.91,50.28) .. (391.91,50.82) .. controls (391.91,51.35) and (389.43,51.79) .. (386.38,51.79) .. controls (383.33,51.79) and (380.85,51.35) .. (380.85,50.82) -- cycle ;
\path  [shading=_4bq82r9c8,_xz6zbha2e,path fading= _b7ie1pk3k ,fading transform={xshift=2}] (395.91,38.84) -- (409.43,25.32) -- (520.63,25.32) -- (520.63,40.92) -- (507.1,54.45) -- (395.91,54.45) -- cycle ; 
 \draw   (395.91,38.84) -- (409.43,25.32) -- (520.63,25.32) -- (520.63,40.92) -- (507.1,54.45) -- (395.91,54.45) -- cycle ; 
 \draw   (520.63,25.32) -- (507.1,38.84) -- (395.91,38.84) ; \draw   (507.1,38.84) -- (507.1,54.45) ;

\draw  [line width=2.25]  (593,315.95) .. controls (593,313.22) and (595.22,311) .. (597.95,311) -- (612.8,311) .. controls (615.53,311) and (617.75,313.22) .. (617.75,315.95) -- (617.75,358.05) .. controls (617.75,360.78) and (615.53,363) .. (612.8,363) -- (597.95,363) .. controls (595.22,363) and (593,360.78) .. (593,358.05) -- cycle ;
\draw  [line width=2.25]  (600.83,313.19) .. controls (600.83,313.02) and (600.97,312.88) .. (601.14,312.88) -- (609.92,312.88) .. controls (610.09,312.88) and (610.23,313.02) .. (610.23,313.19) -- (610.23,313.19) .. controls (610.23,313.37) and (610.09,313.51) .. (609.92,313.51) -- (601.14,313.51) .. controls (600.97,313.51) and (600.83,313.37) .. (600.83,313.19) -- cycle ;
\draw  [line width=1.5]  (600.83,358.3) .. controls (600.83,358.13) and (600.97,357.99) .. (601.14,357.99) -- (609.92,357.99) .. controls (610.09,357.99) and (610.23,358.13) .. (610.23,358.3) -- (610.23,358.3) .. controls (610.23,358.47) and (610.09,358.61) .. (609.92,358.61) -- (601.14,358.61) .. controls (600.97,358.61) and (600.83,358.47) .. (600.83,358.3) -- cycle ;

\draw  [draw opacity=0][shading=_kymnuo23h,_ltfwf60ws] (457.7,208.51) .. controls (499.7,239.51) and (585.5,299) .. (597.95,311) .. controls (610.4,323) and (439.7,199.51) .. (427.7,188.51) .. controls (415.7,177.51) and (415.7,177.51) .. (457.7,208.51) -- cycle ;
\draw  [color={rgb, 255:red, 41; green, 92; blue, 153 }  ,draw opacity=1 ][line width=3]  (258,342) -- (328,342) -- (328,421) -- (258,421) -- cycle ;
\draw  [color={rgb, 255:red, 41; green, 92; blue, 153 }  ,draw opacity=1 ][fill={rgb, 255:red, 188; green, 214; blue, 245 }  ,fill opacity=1 ][line width=3]  (272,356) -- (320.5,356) -- (320.5,370) -- (272,370) -- cycle ;
\draw  [color={rgb, 255:red, 41; green, 92; blue, 153 }  ,draw opacity=1 ][fill={rgb, 255:red, 188; green, 214; blue, 245 }  ,fill opacity=1 ][line width=3]  (272,380) -- (320.5,380) -- (320.5,394) -- (272,394) -- cycle ;
\draw  [color={rgb, 255:red, 41; green, 92; blue, 153 }  ,draw opacity=1 ][fill={rgb, 255:red, 255; green, 255; blue, 255 }  ,fill opacity=1 ][line width=3]  (304,286) -- (374,286) -- (374,421) -- (304,421) -- cycle ;
\draw  [color={rgb, 255:red, 41; green, 92; blue, 153 }  ,draw opacity=1 ][fill={rgb, 255:red, 188; green, 214; blue, 245 }  ,fill opacity=1 ][line width=3]  (320.5,384) -- (338.5,384) -- (338.5,420) -- (320.5,420) -- cycle ;
\draw  [color={rgb, 255:red, 41; green, 92; blue, 153 }  ,draw opacity=1 ][fill={rgb, 255:red, 188; green, 214; blue, 245 }  ,fill opacity=1 ][line width=3]  (339.5,384) -- (356.5,384) -- (356.5,420) -- (339.5,420) -- cycle ;
\draw  [color={rgb, 255:red, 41; green, 92; blue, 153 }  ,draw opacity=1 ][fill={rgb, 255:red, 188; green, 214; blue, 245 }  ,fill opacity=1 ][line width=3]  (315,304) -- (329.5,304) -- (329.5,317) -- (315,317) -- cycle ;
\draw  [color={rgb, 255:red, 41; green, 92; blue, 153 }  ,draw opacity=1 ][fill={rgb, 255:red, 188; green, 214; blue, 245 }  ,fill opacity=1 ][line width=3]  (347,304) -- (361.5,304) -- (361.5,317) -- (347,317) -- cycle ;
\draw  [color={rgb, 255:red, 41; green, 92; blue, 153 }  ,draw opacity=1 ][fill={rgb, 255:red, 188; green, 214; blue, 245 }  ,fill opacity=1 ][line width=3]  (315,328) -- (329.5,328) -- (329.5,341) -- (315,341) -- cycle ;
\draw  [color={rgb, 255:red, 41; green, 92; blue, 153 }  ,draw opacity=1 ][fill={rgb, 255:red, 188; green, 214; blue, 245 }  ,fill opacity=1 ][line width=3]  (347,328) -- (361.5,328) -- (361.5,341) -- (347,341) -- cycle ;
\draw  [color={rgb, 255:red, 41; green, 92; blue, 153 }  ,draw opacity=1 ][fill={rgb, 255:red, 188; green, 214; blue, 245 }  ,fill opacity=1 ][line width=3]  (315,353) -- (329.5,353) -- (329.5,366) -- (315,366) -- cycle ;
\draw  [color={rgb, 255:red, 41; green, 92; blue, 153 }  ,draw opacity=1 ][fill={rgb, 255:red, 188; green, 214; blue, 245 }  ,fill opacity=1 ][line width=3]  (347,353) -- (361.5,353) -- (361.5,366) -- (347,366) -- cycle ;
\draw  [color={rgb, 255:red, 41; green, 92; blue, 153 }  ,draw opacity=1 ][fill={rgb, 255:red, 47; green, 125; blue, 218 }  ,fill opacity=1 ][line width=3]  (309.5,277) -- (367.5,277) -- (367.5,286) -- (309.5,286) -- cycle ;
\draw  [dash pattern={on 4.5pt off 4.5pt}]  (56.63,333.75) -- (302.5,325) ;
\draw  [dash pattern={on 4.5pt off 4.5pt}]  (373.5,321) -- (596.13,310.75) ;
\draw  [draw opacity=0][shading=_m82snt6qh,_vlh58s86n] (456.5,409) .. controls (422.5,408) and (367.5,414) .. (398.5,418) .. controls (429.5,422) and (482.5,419) .. (517.5,418) .. controls (552.5,417) and (490.5,410) .. (456.5,409) -- cycle ;
\draw  [draw opacity=0][shading=_0f8c73vru,_7d6bp3eak] (55,336.5) .. controls (10,362.5) and (68.5,321) .. (55,336.5) .. controls (41.5,352) and (334.5,200) .. (337.5,186) .. controls (340.5,172) and (100,310.5) .. (55,336.5) -- cycle ;
\draw  [draw opacity=0][shading=_in1rcz224,_skk7kfusv] (349.5,186) .. controls (352.5,189) and (341.5,185) .. (337.5,186) .. controls (333.5,187) and (578.5,313) .. (593,315.95) .. controls (607.5,318.9) and (346.5,183) .. (349.5,186) -- cycle ;
\draw  [draw opacity=0][shading=_0rccx2sqs,_660gnrfta] (390.5,446) .. controls (391.5,440) and (399.5,441) .. (415.5,442) .. controls (431.5,443) and (508.5,441) .. (523.5,446) .. controls (538.5,451) and (389.5,452) .. (390.5,446) -- cycle ;
\draw  [dash pattern={on 4.5pt off 4.5pt}]  (497.5,100) -- (451.99,104.79) ;
\draw [shift={(450,105)}, rotate = 353.99] [color={rgb, 255:red, 0; green, 0; blue, 0 }  ][line width=0.75]    (10.93,-3.29) .. controls (6.95,-1.4) and (3.31,-0.3) .. (0,0) .. controls (3.31,0.3) and (6.95,1.4) .. (10.93,3.29)   ;

\draw (98.2,108.5) node   [align=left] {\begin{minipage}[lt]{95.68pt}\setlength\topsep{0pt}
\begin{center}
Active (transmitting) \\element
\end{center}

\end{minipage}};
\draw (69,357.25) node [anchor=north west][inner sep=0.75pt]  [font=\large] [align=left] {{\tiny Radio resource units}};
\draw (76.25,339.25) node [anchor=north west][inner sep=0.75pt]  [font=\large] [align=left] {{\tiny Antenna}};
\draw (453.66,45.28) node  [color={rgb, 255:red, 255; green, 255; blue, 255 }  ,opacity=1 ] [align=left] {Microcontroller};
\draw (332.96,50.67) node  [color={rgb, 255:red, 255; green, 255; blue, 255 }  ,opacity=1 ] [align=left] {\textcolor[rgb]{0.82,0.01,0.11}{Receiver}};
\draw (35.01,293.5) node   [align=left] {BS};
\draw (223.01,71.5) node   [align=left] {SORIS};
\draw (126.2,229.5) node  [rotate=-309.74] [align=left] {Downlink};
\draw (606.01,378.5) node   [align=left] {UE};
\draw (524.2,241.5) node  [rotate=-36.54] [align=left] {Uplink};
\draw (329.02,440.5) node   [align=left] {Blocker};
\draw (530,405) node [anchor=north west][inner sep=0.75pt]   [align=left] {Channel estimation phase};
\draw (524,437) node [anchor=north west][inner sep=0.75pt]   [align=left] {Information transmission phase};
\draw (536.2,100.5) node   [align=left] {\begin{minipage}[lt]{48.08pt}\setlength\topsep{0pt}
\begin{center}
Reflecting\\element
\end{center}

\end{minipage}};
\draw (194.75,307.85) node [anchor=north west][inner sep=0.75pt]  [rotate=-358.24] [align=left] {Blocked};

\end{tikzpicture}

%% file: commProt.tex
\tikzset{every picture/.style={line width=0.75pt}} 

\begin{tikzpicture}[x=0.75pt,y=0.75pt,yscale=-1,xscale=1]

\draw  [fill={rgb, 255:red, 238; green, 171; blue, 171 }  ,fill opacity=0.58 ][line width=2.25]  (27.95,222) -- (352,222) -- (352,273) -- (27.95,273) -- cycle ;
\draw  [line width=3]  (739,82) -- (1102,82) -- (1102,139) -- (739,139) -- cycle ;
\draw  [line width=3]  (468,82) -- (602.5,82) -- (602.5,139) -- (468,139) -- cycle ;
\draw [line width=1.5]    (468.5,71.5) -- (739.5,71.5) ;
\draw [shift={(742.5,71.5)}, rotate = 180] [color={rgb, 255:red, 0; green, 0; blue, 0 }  ][line width=1.5]    (14.21,-4.28) .. controls (9.04,-1.82) and (4.3,-0.39) .. (0,0) .. controls (4.3,0.39) and (9.04,1.82) .. (14.21,4.28)   ;
\draw [shift={(465.5,71.5)}, rotate = 0] [color={rgb, 255:red, 0; green, 0; blue, 0 }  ][line width=1.5]    (14.21,-4.28) .. controls (9.04,-1.82) and (4.3,-0.39) .. (0,0) .. controls (4.3,0.39) and (9.04,1.82) .. (14.21,4.28)   ;
\draw [line width=1.5]    (465.5,35.5) -- (1097,35.5) ;
\draw [shift={(1100,35.5)}, rotate = 180] [color={rgb, 255:red, 0; green, 0; blue, 0 }  ][line width=1.5]    (14.21,-4.28) .. controls (9.04,-1.82) and (4.3,-0.39) .. (0,0) .. controls (4.3,0.39) and (9.04,1.82) .. (14.21,4.28)   ;
\draw [shift={(462.5,35.5)}, rotate = 0] [color={rgb, 255:red, 0; green, 0; blue, 0 }  ][line width=1.5]    (14.21,-4.28) .. controls (9.04,-1.82) and (4.3,-0.39) .. (0,0) .. controls (4.3,0.39) and (9.04,1.82) .. (14.21,4.28)   ;
\draw  [fill={rgb, 255:red, 238; green, 171; blue, 171 }  ,fill opacity=0.58 ][line width=1.5]  (27.95,222) -- (75.87,222) -- (75.87,273) -- (27.95,273) -- cycle ;
\draw  [fill={rgb, 255:red, 238; green, 171; blue, 171 }  ,fill opacity=0.58 ][line width=1.5]  (75.59,222) -- (123.52,222) -- (123.52,273) -- (75.59,273) -- cycle ;
\draw  [fill={rgb, 255:red, 238; green, 171; blue, 171 }  ,fill opacity=0.58 ][line width=1.5]  (185.41,221) -- (233.33,221) -- (233.33,272) -- (185.41,272) -- cycle ;
\draw  [fill={rgb, 255:red, 238; green, 171; blue, 171 }  ,fill opacity=0.58 ][line width=1.5]  (304.08,222) -- (352,222) -- (352,273) -- (304.08,273) -- cycle ;
\draw  [fill={rgb, 255:red, 238; green, 171; blue, 171 }  ,fill opacity=0.58 ][line width=1.5]  (468,82) -- (482,82) -- (482,139) -- (468,139) -- cycle ;
\draw [line width=1.5]  [dash pattern={on 1.69pt off 2.76pt}]  (468,139) -- (27.95,222) ;
\draw [line width=1.5]  [dash pattern={on 1.69pt off 2.76pt}]  (482,139) -- (352,222) ;
\draw  [fill={rgb, 255:red, 238; green, 171; blue, 171 }  ,fill opacity=0.58 ][line width=1.5]  (588.5,82) -- (602.5,82) -- (602.5,139) -- (588.5,139) -- cycle ;
\draw  [fill={rgb, 255:red, 238; green, 171; blue, 171 }  ,fill opacity=0.58 ][line width=2.25]  (380.95,222) -- (705,222) -- (705,273) -- (380.95,273) -- cycle ;
\draw  [fill={rgb, 255:red, 238; green, 171; blue, 171 }  ,fill opacity=0.58 ][line width=1.5]  (380.95,222) -- (428.87,222) -- (428.87,273) -- (380.95,273) -- cycle ;
\draw  [fill={rgb, 255:red, 238; green, 171; blue, 171 }  ,fill opacity=0.58 ][line width=1.5]  (428.59,222) -- (476.52,222) -- (476.52,273) -- (428.59,273) -- cycle ;
\draw  [fill={rgb, 255:red, 238; green, 171; blue, 171 }  ,fill opacity=0.58 ][line width=1.5]  (538.41,221) -- (586.33,221) -- (586.33,272) -- (538.41,272) -- cycle ;
\draw  [fill={rgb, 255:red, 238; green, 171; blue, 171 }  ,fill opacity=0.58 ][line width=1.5]  (657.08,222) -- (705,222) -- (705,273) -- (657.08,273) -- cycle ;
\draw [line width=1.5]  [dash pattern={on 1.69pt off 2.76pt}]  (588.5,139) -- (380.95,222) ;
\draw [line width=1.5]  [dash pattern={on 1.69pt off 2.76pt}]  (602.5,139) -- (705,222) ;
\draw  [line width=3]  (604,82) -- (738.5,82) -- (738.5,139) -- (604,139) -- cycle ;
\draw  [fill={rgb, 255:red, 213; green, 233; blue, 243 }  ,fill opacity=0.58 ][line width=1.5]  (604,82) -- (618,82) -- (618,139) -- (604,139) -- cycle ;
\draw  [fill={rgb, 255:red, 213; green, 233; blue, 243 }  ,fill opacity=1 ][line width=1.5]  (724.5,82) -- (738.5,82) -- (738.5,139) -- (724.5,139) -- cycle ;
\draw  [fill={rgb, 255:red, 213; green, 233; blue, 243 }  ,fill opacity=1 ][line width=2.25]  (705.95,222) -- (1030,222) -- (1030,273) -- (705.95,273) -- cycle ;
\draw  [fill={rgb, 255:red, 213; green, 233; blue, 243 }  ,fill opacity=1 ][line width=1.5]  (705.95,222) -- (753.87,222) -- (753.87,273) -- (705.95,273) -- cycle ;
\draw  [fill={rgb, 255:red, 213; green, 233; blue, 243 }  ,fill opacity=1 ][line width=1.5]  (753.59,222) -- (801.52,222) -- (801.52,273) -- (753.59,273) -- cycle ;
\draw  [fill={rgb, 255:red, 213; green, 233; blue, 243 }  ,fill opacity=1 ][line width=1.5]  (863.41,221) -- (911.33,221) -- (911.33,272) -- (863.41,272) -- cycle ;
\draw  [fill={rgb, 255:red, 213; green, 233; blue, 243 }  ,fill opacity=1 ][line width=1.5]  (982.08,222) -- (1030,222) -- (1030,273) -- (982.08,273) -- cycle ;
\draw [line width=1.5]  [dash pattern={on 1.69pt off 2.76pt}]  (618,139) -- (1030,222) ;
\draw  [fill={rgb, 255:red, 213; green, 233; blue, 243 }  ,fill opacity=1 ][line width=2.25]  (1063.95,223) -- (1388,223) -- (1388,274) -- (1063.95,274) -- cycle ;
\draw  [fill={rgb, 255:red, 213; green, 233; blue, 243 }  ,fill opacity=1 ][line width=1.5]  (1063.95,223) -- (1111.87,223) -- (1111.87,274) -- (1063.95,274) -- cycle ;
\draw  [fill={rgb, 255:red, 213; green, 233; blue, 243 }  ,fill opacity=1 ][line width=1.5]  (1111.59,223) -- (1159.52,223) -- (1159.52,274) -- (1111.59,274) -- cycle ;
\draw  [fill={rgb, 255:red, 213; green, 233; blue, 243 }  ,fill opacity=1 ][line width=1.5]  (1221.41,222) -- (1269.33,222) -- (1269.33,273) -- (1221.41,273) -- cycle ;
\draw  [fill={rgb, 255:red, 213; green, 233; blue, 243 }  ,fill opacity=1 ][line width=1.5]  (1340.08,223) -- (1388,223) -- (1388,274) -- (1340.08,274) -- cycle ;
\draw [line width=1.5]  [dash pattern={on 1.69pt off 2.76pt}]  (739,139) -- (1388,223) ;
\draw [line width=1.5]  [dash pattern={on 1.69pt off 2.76pt}]  (724.5,139) -- (1063.95,223) ;

\draw (535.25,110.5) node   [align=left] {Downlink};
\draw (671.25,110.5) node   [align=left] {Uplink};
\draw (920.5,110.5) node   [align=left] {Data transmission};
\draw (611.21,55.5) node   [align=left] {Channel estimation};
\draw (743.21,19.5) node   [align=left] {Transmission cycle};
\draw (51.91,247.5) node    {$x_{1}( 1)$};
\draw (100.55,247.5) node    {$x_{1}( 2)$};
\draw (328.04,247.5) node    {$x_{1}( L)$};
\draw (209.37,246.5) node    {$x_{1}( l)$};
\draw (153.59,249) node    {$\cdots $};
\draw (268.59,248) node    {$\cdots $};
\draw (299.45,199.5) node   [align=left] {\begin{minipage}[lt]{78.13pt}\setlength\topsep{0pt}
\begin{center}
\footnotesize{Active element $1$}\\\footnotesize{of set $\displaystyle \mathcal{S}$}
\end{center}

\end{minipage}};
\draw (404.91,247.5) node    {$x_{N_{d}}( 1)$};
\draw (453.55,247.5) node    {$x_{N_{d}}( 2)$};
\draw (681.04,247.5) node    {$x_{N_{d}}( L)$};
\draw (562.37,246.5) node    {$x_{N_{d}}( l)$};
\draw (506.59,249) node    {$\cdots $};
\draw (621.59,248) node    {$\cdots $};
\draw (566.45,197.5) node   [align=left] {\begin{minipage}[lt]{86.94pt}\setlength\topsep{0pt}
\begin{center}
\footnotesize{Active element $\displaystyle N_{f}$}\\\footnotesize{of set $\displaystyle \mathcal{S}$}
\end{center}

\end{minipage}};
\draw (364.59,249) node    {$\cdots $};
\draw (729.91,247.5) node    {$y_{1}( 1)$};
\draw (778.55,247.5) node    {$y_{1}( 2)$};
\draw (1006.04,247.5) node    {$y_{1}( L)$};
\draw (887.37,246.5) node    {$y_{1}( l)$};
\draw (831.59,249) node    {$\cdots $};
\draw (946.59,248) node    {$\cdots $};
\draw (746.45,199.5) node   [align=left] {\begin{minipage}[lt]{78.13pt}\setlength\topsep{0pt}
\begin{center}
\footnotesize{Active element 1}\\\footnotesize{of set $\displaystyle \mathcal{S}$}
\end{center}

\end{minipage}};
\draw (1087.91,248.5) node    {$y_{N_{f}}( 1)$};
\draw (1136.55,248.5) node    {$y_{N_{f}}( 2)$};
\draw (1364.04,248.5) node    {$y_{N_{f}}( L)$};
\draw (1245.37,247.5) node    {$y_{N_{f}}( l)$};
\draw (1189.59,250) node    {$\cdots $};
\draw (1304.59,249) node    {$\cdots $};
\draw (984.45,190.5) node  [rotate=-13.37] [align=left] {\begin{minipage}[lt]{125.44pt}\setlength\topsep{0pt}
\begin{center}
\footnotesize{Active element $\displaystyle N_{f}$ of set $\displaystyle \mathcal{S}$}
\end{center}

\end{minipage}};
\draw (1047.59,247) node    {$\cdots $};

\end{tikzpicture}

%% file: Abs.tex
\begin{tikzpicture}

\begin{axis} [
    x tick label style={/pgf/number format/1000 sep},
    xmode=log,
    log base x={2},
    ylabel={Average Mean Square Error},
    xlabel={Number of active elements},
    enlargelimits=0.04,
    legend style={at={(0.85,1)},
    anchor=north,legend columns=1},
    ybar interval=0.7,
    grid style=dashed,
    ]

\addplot
	coordinates {
        (4, 0.0018823720852378756 )
        (8, 0.001725902515463531 )
        (16, 0.0014446598733775318 )
        (32, 0.0010912127199117095 )
        (64, 0)
	};

\addplot
	coordinates {
        (4, 0.0017458500550128519 )
        (8, 0.0013187366398051382 )
        (16, 0.0005030800512759015 )
        (32, 0.00015915788851998515 )
        (64, 0 )
	};

\addplot
	coordinates {
        (4, 0.0013214071723632514 )
        (8, 0.00031661799555877224 )
        (16, 5.459183357743313e-05 )
        (32, 3.990106588389608e-05 )
        (64, 0 )
	};

\addplot
	coordinates {
        (4, 0.00022074437511037103 )
        (8, 4.454487989278277e-05 )
        (16, 1.381238258659323e-05 )
        (32, 1.3024621548538562e-05 )
        (64, 0 )
	};

\legend{$\lambda/2$, $\lambda/4$, $\lambda/8$, $\lambda/16$}

\end{axis}

\end{tikzpicture}

%% file: Angles.tex
\begin{tikzpicture}

\begin{axis} [
    x tick label style={/pgf/number format/1000 sep},
    xmode=log,
    log base x={2},
    ylabel={Average Mean Square Error},
    xlabel={Number of active elements},
    enlargelimits=0.04,
    legend style={at={(0.8,0.95)},
    anchor=north,legend columns=1},
    ybar interval=0.7,
    grid style=dashed,
]

\addplot
	coordinates {
        (4, 0.018949264325201512 )
        (8, 0.01735663093626499 )
        (16, 0.014466101611033082 )
        (32, 0.009862748105078936 )
        (64, 0 )
	};

\addplot
	coordinates {
        (4, 0.01771988669410348 )
        (8, 0.013510455684736371 )
        (16, 0.005084953736513853 )
        (32, 0.0011359916278161108 )
        (64, 0 )
	};

\addplot
	coordinates {
        (4, 0.013098408747464419 )
        (8, 0.0033656804263591765 )
        (16, 0.000433110143058002 )
        (32, 0.0001608555890925345 )
        (64, 0 )
	};

\addplot
	coordinates {
        (4, 0.0024367523868568243 )
        (8, 0.00043247140856692565 )
        (16, 7.635793612280395e-05 )
        (32, 7.288894013981917e-05 )
        (64, 0 )
	};

\legend{$\lambda/2$, $\lambda/4$, $\lambda/8$, $\lambda/16$}

\end{axis}
\end{tikzpicture}

%% file: 16x16n32x32Angles.tex
\begin{tikzpicture}

\begin{axis} [
    x tick label style={/pgf/number format/1000 sep},
    xmode=log,
    log base x={2},
    ylabel={Average Mean Square Error},
    xlabel={Number of active elements},
    enlargelimits=0.04,
    legend style={at={(1.15,0.95)},
    anchor=north,legend columns=1},
    ybar interval=0.7,
    grid style=dashed,
    ]

\addplot
	coordinates {
        (4, 0.015367710329592228 )
        (8, 0.006540850354358554 )
        (16, 0.0013560055318521335 )
        (32, 0 )
	};
\addlegendentry{\hspace{-2.5em} 256 RIS phases}

\addplot
	coordinates {
        (4, 0.020196729972958564 )
        (8, 0.018207472935318946 )
        (16, 0.01665909039787948 )
        (32, 0)
	};
\addlegendentry{\hspace{-2em} 1024 RIS phases}

\addplot
	coordinates {
        (4, 0.0008314227144001052 )
        (8, 0.0003882126527605578 )
        (16, 0.00016494984085511534 )
        (32, 0 )
	};
\addlegendentry{\hspace{-0.5em}256 RIS magnitude}

\addplot
	coordinates {
        (4, 0.0006022451823810115 )
        (8, 0.0005940863455180078 )
        (16, 0.0005786240272573196 )
        (32, 0)
	};
\addlegendentry{1024 RIS magnitude}
        

\end{axis}

\end{tikzpicture}

%% file: 4active_Abs.tex
\begin{tikzpicture}

\begin{axis} [
    x tick label style={/pgf/number format/1000 sep},
    ylabel={Average Mean Square Error},
    xlabel={Set number},
    enlargelimits=0.04,
    legend style={at={(0.8,0.8)},
    anchor=north,legend columns=1},
    ybar interval=0.7,
    grid style=dashed,
]

\addplot
	coordinates {
        (1, 0.0018824843736365438 )
        (2, 0.0018644213327206672 )
        (3, 0.0018507063202559948 )
        (4, 0.0018605719425249844 )
        (5, 1E-5 )
	};

\addplot
	coordinates {
        (1, 0.001747967597329989 )
        (2, 0.0016486710193566979 )
        (3, 0.0015476884669624268 )
        (4, 0.001595928044989705 )
        (5, 1E-5 )
	};

\addplot
	coordinates {
        (1, 0.0013231559249106794 )
        (2, 0.0009538962930673733 )
        (3, 0.0006656177341938019 )
        (4, 0.0007498791394755244 )
        (5, 1E-5 )
	};

\addplot
	coordinates {
        (1, 0.00022145711845951156 )
        (2, 0.00011373231020115781 )
        (3, 0.00010366085749410558 )
        (4, 6.396327997208573e-05 )
        (5, 1E-5 )
	};

\legend{$\lambda/2$, $\lambda/4$, $\lambda/8$, $\lambda/16$}

\end{axis}
\end{tikzpicture}

%% file: 4active_Angle.tex
\begin{tikzpicture}

\begin{axis} [
    x tick label style={/pgf/number format/1000 sep},
    ylabel={Average Mean Square Error},
    xlabel={Set number},
    enlargelimits=0.04,
    legend style={at={(0.8,0.8)},
    anchor=north,legend columns=1},
    ybar interval=0.7,
    grid style=dashed,
]

\addplot
	coordinates {
        (1, 0.018952720891684295 )
        (2, 0.018839422296732664 )
        (3, 0.0187125145830214 )
        (4, 0.018803517799824476 )
        (5, 1E-5)
	};

\addplot
	coordinates {
        (1, 0.017720790039747955 )
        (2, 0.01689134180545807 )
        (3, 0.01582559697329998 )
        (4, 0.016276569925248622 )
        (5, 1E-5)
	};

\addplot
	coordinates {
        (1, 0.013102093869820237 )
        (2, 0.00970045301131904 )
        (3, 0.0070847814483568075 )
        (4, 0.007711599944159388 )
        (5, 1E-5)
	};

\addplot
	coordinates {
        (1, 0.0024256527959369122 )
        (2, 0.0012963250139728189 )
        (3, 0.0011691884580068291 )
        (4, 0.0007549328019376844 )
        (5, 1E-5)
	};

\legend{$\lambda/2$, $\lambda/4$, $\lambda/8$, $\lambda/16$}

\end{axis}
\end{tikzpicture}

%% file: 8active_Abs_new1.tex
\begin{tikzpicture}

\begin{axis} [
    x tick label style={/pgf/number format/1000 sep},
    ylabel={Average Mean Square Error},
    xlabel={Set number},
    enlargelimits=0.04,
    legend style={at={(0.9, 1.01)},
    anchor=north,legend columns=1},
    ybar interval=0.7,
    grid style=dashed,
]

\addplot
	coordinates {
        (1, 0.0017242090648505837 )
        (2, 0.0017373488482553513 )
        (3, 0.0017185048863757402 )
        (4, 0.0017066928395070135 )
        (5, 1E-5 )
	};

\addplot
	coordinates {
        (1, 0.0013079165306407957 )
        (2, 0.0013836577499751002 )
        (3, 0.0011638722522184252 )
        (4, 0.0011432043032255023 )
        (5, 1E-5 )
	};

\addplot
	coordinates {
        (1, 0.0003647852034191601 )
        (2, 0.0003686389292124659 )
        (3, 0.00011600347905186936 )
        (4, 0.0003210567889618687 )
        (5, 1E-5 )
	};

\addplot
	coordinates {
        (1, 5.078474805486621e-05 )
        (2, 1.9635671301330148e-05 )
        (3, 1.718780860073821e-05 )
        (4, 4.533456340141129e-05 )
        (5, 1E-5 )
	};

\legend{$\lambda/2$, $\lambda/4$, $\lambda/8$, $\lambda/16$}

\end{axis}
\end{tikzpicture}

%% file: 8active_Angle_new1.tex
\begin{tikzpicture}

\begin{axis} [
    x tick label style={/pgf/number format/1000 sep},
    ylabel={Average Mean Square Error},
    xlabel={Set number},
    enlargelimits=0.04,
    legend style={at={(0.9,1.03)},
    anchor=north,legend columns=1},
    ybar interval=0.7,
    grid style=dashed,
]

\addplot
	coordinates {
        (1, 0.017351469676941633 )
        (2, 0.017491284199059008 )
        (3, 0.017225714474916456 )
        (4, 0.017256574928760527 )
        (5, 1E-5)
	};

\addplot
	coordinates {
        (1, 0.013376931212842464 )
        (2, 0.014214874487370252 )
        (3, 0.011916736783459782 )
        (4, 0.01182964381761849 )
        (5, 1E-5)
	};

\addplot
	coordinates {
        (1, 0.003909861342981458 )
        (2, 0.0038716269307769837 )
        (3, 0.0013800577737856656 )
        (4, 0.003442772435955703 )
        (5, 1E-5)
	};

\addplot
	coordinates {
        (1, 0.00047850851027760657 )
        (2, 0.00020674811225035228 )
        (3, 0.00011627199353824835 )
        (4, 0.0004462139229872264 )
        (5, 1E-5)
	};

\legend{$\lambda/2$, $\lambda/4$, $\lambda/8$, $\lambda/16$}

\end{axis}
\end{tikzpicture}

%% file: Angles_compare.tex
\begin{tikzpicture}

\begin{axis} [
    x tick label style={/pgf/number format/1000 sep},
    xmode=log,
    log base x={2},
    ylabel={Average Mean Square Error},
    xlabel={Number of active elements},
    enlargelimits=0.04,
    legend style={at={(0.9,0.95)},
    anchor=north,legend columns=1},
    ybar interval=0.7,
    grid style=dashed,
]

\addplot
	coordinates {
        (4, 0.014526071436703205 )
        (8, 0.013856257433071733 )
        (16, 0.012254816992208362 )
        (32, 0.01276228935457766 )
        (64, 0 )
	};

\addplot
	coordinates {
        (4, 0.003248042066115886 )
        (8, 0.0010221643882687204 )
        (16, 0.0006500356702599674 )
        (32, 0.00048717811034293846 )
        (64, 0 )
	};

\addplot
	coordinates {
        (4, 0.0024367523868568243 )
        (8, 0.00043247140856692565 )
        (16, 7.635793612280395e-05 )
        (32, 7.288894013981917e-05 )
        (64, 0 )
	};

\legend{CNN, LI, RNN}

\end{axis}
\end{tikzpicture}

%% file: Angles_predict.tex
    \begin{tikzpicture}[x=0.55pt,y=0.55pt,yscale=1,xscale=1]
    \begin{axis}[
        xlabel={$\sigma$},
        ylabel={AMSE},
        ymode=log,
        xticklabel style={
        /pgf/number format/precision=2 
         },
        xtick={0, 0.01, 0.02, 0.03, 0.04, 0.05, 0.06, 0.07, 0.08, 0.09, 0.1},
        ymax=0.1, 
        grid=major,
        width=14cm, 
        height=10cm, 
        legend pos=outer north east, 
        legend style={at={(0.8,0.5)},anchor=north}, 
        axis lines=left,
        smooth, 
    ]

    \addplot[
        red,
        thick,
        mark=o, 
        mark size=2pt,
    ]
    coordinates {
        (0.0, 1.325221819570288e-05)
        (0.01, 8.501578849973157e-05)
        (0.02, 0.00030063805752433836)
        (0.03, 0.0006537071894854307)
        (0.04, 0.0011462144320830703)
        (0.05, 0.001774532487615943)
        (0.06, 0.0025252767372876406)
        (0.07, 0.003399364184588194)
        (0.08, 0.004385494161397219)
        (0.09, 0.005460659042000771)
        (0.1,  0.006645588669925928)
    };
    \addlegendentry{32 active elements}

    \addplot[
        blue,
        thick,
        mark=square*, 
        mark size=2pt,
    ]
    coordinates {
        (0.0, 2.5289909899584018e-05)
        (0.01, 0.0002099062839988619)
        (0.02, 0.0007606277358718216)
        (0.03, 0.0016590976156294346)
        (0.04, 0.0028971966821700335)
        (0.05, 0.0044143302366137505)
        (0.06, 0.006193515844643116)
        (0.07, 0.008233225904405117)
        (0.08, 0.01042971108108759)
        (0.09, 0.012827121652662754)
        (0.1,  0.015322973020374775)
    };
    \addlegendentry{16 active elements}

    \addplot[
        green!60!black,
        thick, 
        mark=diamond*, 
        mark size=2pt,
    ]
    coordinates {
        (0.0, 0.0003673634782899171)
        (0.01, 0.00048200617311522365)
        (0.02, 0.0008206046768464148)
        (0.03, 0.0013815833954140544)
        (0.04, 0.0021598837338387966)
        (0.05, 0.00313136400654912)
        (0.06, 0.004287962801754475)
        (0.07, 0.005618819035589695)
        (0.08, 0.007126144133508205)
        (0.09, 0.008737731724977493)
        (0.1,  0.010508282110095024)
    };
    \addlegendentry{8 active elements}

    \addplot[
        magenta,
        thick, 
        mark=triangle*, 
        mark size=2pt,
    ]
    coordinates {
        (0.0, 0.002318363869562745)
        (0.01, 0.0024784980341792107)
        (0.02, 0.00294507946819067)
        (0.03, 0.0037191978190094233)
        (0.04, 0.004756464157253504)
        (0.05, 0.00605334946885705)
        (0.06, 0.007556437980383635)
        (0.07, 0.009272105991840363)
        (0.08, 0.01112463977187872)
        (0.09, 0.01311424095183611)
        (0.1,  0.015244334004819393)
    };
    \addlegendentry{4 active elements}

    \end{axis}
    \end{tikzpicture}

%% file: abs_predict.tex
    \begin{tikzpicture}
    \begin{axis}[
        xlabel={$\sigma$},
        ylabel={AMSE},
        ymode=log,
        xticklabel style={
                        /pgf/number format/fixed,
                        /pgf/number format/precision=2 
         },
        xtick={0, 0.01, 0.02, 0.03, 0.04, 0.05, 0.06, 0.07, 0.08, 0.09, 0.1},
        ymax=0.1,  
        grid=major,
        width=14cm, 
        height=10cm, 
        legend pos=outer north east, 
        legend style={at={(0.3,0.95)},anchor=north}, 
        axis lines=left,
        smooth, 
    ]

    \addplot[
        red,
        thick,
        mark=o, 
        mark size=2pt,
    ]
    coordinates {
        (0.0, 1.486068072154012e-06)
        (0.01, 0.0001397417508997023)
        (0.02, 0.0005495315417647362)
        (0.03, 0.0012130410177633166)
        (0.04, 0.002114303642883897)
        (0.05, 0.003220402402803302)
        (0.06, 0.0044984351843595505)
        (0.07, 0.00594156701117754)
        (0.08, 0.007492242380976677)
        (0.09, 0.0091165816411376)
        (0.1, 0.010822762735188007)
    };
    \addlegendentry{32 active elements}

    \addplot[
        blue,
        thick,
        mark=square*, 
        mark size=2pt,
    ]
    coordinates {
        ((0.0, 2.048776423180243e-06)
        (0.01, 0.00026576381060294807)
        (0.02, 0.0010411491384729743)
        (0.03, 0.0022840111050754786)
        (0.04, 0.003932202234864235)
        (0.05, 0.005939014256000519)
        (0.06, 0.008170712739229202)
        (0.07, 0.010634901002049446)
        (0.08, 0.013180767185986042)
        (0.09, 0.015815379098057747)
        (0.1, 0.01857302151620388)
    };
    \addlegendentry{16 active elements}

    \addplot[
        green!60!black,
        thick, 
        mark=diamond*, 
        mark size=2pt,
    ]
    coordinates {
        (0.0,  3.494614065857604e-05)
        (0.01, 0.00021566609211731702)
        (0.02, 0.0007501895888708532)
        (0.03, 0.00161333242431283)
        (0.04, 0.002790778176859021)
        (0.05, 0.004218216519802809)
        (0.06, 0.00585391279309988)
        (0.07, 0.007652604021131992)
        (0.08, 0.009644811041653156)
        (0.09, 0.011732567101716995)
        (0.1,  0.013858674094080925)
    };
    \addlegendentry{8 active elements}

    \addplot[
        magenta,
        thick, 
        mark=triangle*, 
        mark size=2pt,
    ]
    coordinates {
        ((0.0, 0.00020402917289175093)
        (0.01, 0.0004467992694117129)
        (0.02, 0.0011589364148676395)
        (0.03, 0.0023056913632899523)
        (0.04, 0.0038288820069283247)
        (0.05, 0.005677275825291872)
        (0.06, 0.007791580632328987)
        (0.07, 0.010073596611618996)
        (0.08, 0.012545507401227951)
        (0.09, 0.01510950829833746)
        (0.1,  0.01770632341504097)
    };
    \addlegendentry{4 active elements}

    \end{axis}
    \end{tikzpicture}

%% file: Bigger_RIS_BER.tex
\begin{tikzpicture}
\begin{axis} [
    x tick label style={/pgf/number format/1000 sep},    
    xmode=log,
    log basis x={2},
    xtick={64,128,256},
    xticklabels={{$8\times8$},
        {$16\times16$},
        {$32\times32$}},
    ylabel={BER},
    xlabel={Metasurface size},
    enlargelimits=0.04,
    legend style={at={(0.85,0.95)},
    anchor=north,legend columns=1},
    ybar interval=0.7,
    grid style=dashed,
]


\addplot
	coordinates {
        (64, 0.1543715)
        (128, 0.0191461)
        (256, 3.8e-05)
        (512, 0 )
	};

\addplot
	coordinates {
        (64, 0.1435497)
        (128,  0.0110584)
        (256, 3.65e-05)
        (512, 0 )
	};

\addplot
	coordinates {
        (64, 0.1315809)
        (128, 0.005295)
        (256, 3.52e-05)
        (512, 0)
	};


 \addplot
	coordinates {
        (64, 0.1309763)
        (128, 0.005752) 
        (256, 2.04e-05) 
        (512, 0 )
	};
\legend{4,8,16, Ideal}

\end{axis}
\end{tikzpicture}